\documentclass[twocolumn,journal]{IEEEtran}
\usepackage[noadjust]{cite}

\ifCLASSINFOpdf
   \usepackage[pdftex]{graphicx}
\else
   \usepackage[dvips]{graphicx}
\fi
\usepackage[colorlinks,
            linkcolor=black,
            urlcolor=red,
            anchorcolor=black,
            citecolor=black
            ]{hyperref}
\usepackage{tabularx}
\usepackage{supertabular}
\usepackage{amsmath,amssymb,amsfonts}
\usepackage{mathrsfs}
\usepackage{multirow}
\usepackage{algorithm}
\usepackage{algpseudocode}
\usepackage{bm}
\usepackage{longtable}
\usepackage{xcolor}
\usepackage{lineno}
\usepackage{tikz}
\usepackage{circuitikz}
\usepackage{booktabs}%

\usepackage{fixltx2e}
\usepackage{url}

\begin{document}
\title{Entanglement-Assisted Concatenated Quantum Codes: Parameters and Asymptotic Performance}
%

\author{Jihao~Fan,~\IEEEmembership{Member,~IEEE},~Wei~Cheng,~\IEEEmembership{Member,~IEEE}, Gaojun Luo,~\IEEEmembership{Member,~IEEE},~Zhou Li,~\IEEEmembership{Member,~IEEE}, and Meng Cao,~\IEEEmembership{Member,~IEEE}
\thanks{J. Fan is with School of Cyber Science and Engineering, Nanjing University of Science and Technology, Nanjing 210094, China, and also with Laboratory for Advanced Computing and Intelligence Engineering, Wuxi 214083, China  (e-mail: jihao.fan@outlook.com)   }
\thanks{W. Cheng is with LTCI, T\'el\'ecom Paris, Institut Polytechnique de Paris, Palaiseau 91120, France  (e-mail: wei.cheng@telecom-paris.com)   }
\thanks{Gaojun Luo is with the School of Mathematics, Nanjing University of Aeronautics and Astronautics, Nanjing 211106,  China (e-mail: gaojun\_luo@nuaa.edu.cn)}
\thanks{Zhou Li   is  with School of Cyber Science and Engineering, Nanjing University of Science and Technology, Nanjing 210094, China, and also with School of Computer, Electronics and Information, Guangxi University, Nanning 530004, China    (e-mail: Zhouli@my.unt.edu)}
\thanks{Meng Cao is  with Beijing Institute of Mathematical Sciences and Applications, Beijing 101408, China    (e-mail: mengcaomath@126.com)}
 }

\maketitle

\begin{abstract}
Entanglement-assisted concatenated quantum codes (EACQCs) are constructed by concatenating two    entanglement-assisted quantum error-correcting codes (EAQECCs). By selecting the inner and outer component codes  carefully, it is able to construct   state-of-the-art EACQCs with parameters better than   previous quantum codes. In this work, we  use  almost maximum-distance-separable (MDS)  codes and $\hbar$-MDS codes as  the outer   codes to construct EACQCs. Because the range of code length of almost MDS and $\hbar$-MDS codes is much more free than that of the commonly used MDS codes.
We   derive  several families of new EACQCs with parameters better than the previously best known EAQECCs and standard quantum error-correcting codes (QECCs) of the same length and net transmissions.  Moreover, we demonstrate that EACQCs are with maximal entanglement if both the inner  and outer component codes are with maximal entanglement.  As a result, we construct three new maximal-entanglement EACQCs     which have optimal parameters. In addition, we present several new maximal-entanglement EACQCs whose minimum distance is only one less than the minimum distance of the optimal codes. In particular,  we propose     two new families of asymptotically good maximal-entanglement EACQCs with explicit constructions  by using entanglement-assisted quantum algebraic geometry  codes  as the outer codes.  At last, we prove that EACQCs can attain the quantum Gilbert-Varshamov bound for EAQECCs asymptotically.
\end{abstract}


%
\IEEEpeerreviewmaketitle


\newtheorem{definitions}{Definition}
\newtheorem{theorems}{Theorem}
\newtheorem{lemmas}{Lemma}
\newtheorem{corollarys}{Corollary}
\newtheorem{examples}{Example}
\newtheorem{propositions}{Proposition}
%
%
%
%

\section{Introduction}
\label{sec1}

\IEEEPARstart{Q}{uantum} error correction (QEC) is   fundamental in the near or future   realizing of practical and large-scale quantum computers and   quantum communications ~\cite{deutsch2020harnessing}. How to construct  good quantum error-correcting  codes (QECCs) has become an important and hot topic both in quantum information theory \cite{dauphinais2024stabilizer,kibe2022holographic}  and in practical quantum systems \cite{PRXQuantum.5.030326}.  The \emph{stabilizer} formalism provides a general framework   for constructing QECCs from classical additive codes satisfying  certain dual-containing constraint \cite{calderbank1998quantum,ketkar2006nonbinary}. However, such dual-containing relationship  cannot be established directly for most classical error-correcting codes and sometimes  largely restricts the construction of QECCs in the view point of coding theory \cite{mackay2004sparse,aly2007quantum,la2011new}.  Excitingly,
the appearance of entanglement-assisted (EA) quantum error correcting codes (EAQECCs) greatly simplified the construction of quantum codes. Through importing some pre-shared entangled states between the sender (Alice) and the receiver (Bob), it is able to construct EAQECCs from any classical linear code without needing to satisfy the dual-containing constraint \cite{brun2006correcting,wilde2008optimal}.


The first EAQECC was derived  by Bowen \cite{bowen2002entanglement}. The general formalism of constructing EAQECCs was first proposed in \cite{brun2006correcting}. In \cite{hsieh2011high,fujiwara2010entanglement,hsieh2009entanglement}, entanglement-assisted quantum low-density parity-check (LDPC) codes were continuously constructed by importing some amount of pre-shared ebits. In \cite{brun2014catalytic}, EAQECCs with catalytic quantum error correction were established. The optimal entanglement formulas for general EAQECCs were given in \cite{wilde2008optimal}. In \cite{fan2016constructions}, entanglement-assisted quantum  maximum-distance-separable (MDS) codes were first constructed and the number of pre-shared entangled states was computed by using the entanglement formulas for Hermitian type EAQECCs. In \cite{pereira2021entanglement}, EAQECCs were constructed from algebraic geometry (AG) codes and asymptotically good entanglement-assisted quantum algebraic geometry (EAQAG) codes surpassing the quantum Gilbert-Varshamov (GV) bound were constructed. Although EAQECCs  show large advantages compared to standard QECCs, an amount  of pre-shared entanglement is needed before the transmission. Therefore EAQECCs with as small as possible pre-shared entanglement are preferable \cite{brun2006correcting,hsieh2011high} since maintaining  a large mount of entanglement is very expensive. On the other hand, entanglement is precious and  useful physical resource. The pre-shared entanglement can not only facilitate the construction  of QECCs but also   enhance
the transmission rate of QECCs. Therefore a particularly interesting type of EAQECCs that exploit maximal entanglement is useful, e.g., to achieve the entanglement-assisted quantum capacity of quantum channels \cite{hsieh2008entanglement,devetak2008resource}. Moreover, it was shown that EAQECCs with maximal entanglement are highly related to   linear complementary dual (LCD) codes which are useful in resisting side-channel attacks and fault
injection attacks \cite{carlet2018linear,luo2018mds}.

In \cite{fan2022entanglement}, a general framework of entanglement-assisted concatenated quantum codes (EACQCs) was proposed.  It is a generalization of the standard concatenated quantum codes (CQCs)  \cite{knill1996concatenated} through exploiting entanglement assistance. Several families of EACQCs with parameters better than the   previously best known QECCs and EAQECCs were derived by using entanglement-assisted quantum MDS (EAQMDS) codes as the outer. In particular, how to construct EACQCs with better parameters than the best known QECCs and EAQECCs   is state-of-the-art in quantum coding theory. It is known that each nondegenerate EAQECC with best known parameters can be derived from a classical best known linear code (BKLC)  directly. Moreover, error degeneracy is a specific phenomenon that only exists in QECCs and it is   conjectured  that \emph{degenerate} EAQECCs (QECCs) may  have more superior  parameters than nondegenerate ones. Therefore constructing (degenerate) EACQCs with better parameters than the best known nondegenerate EAQECCs can exhibit   supremacy of quantum information theory to classical information theory.

 In this work, we use almost MDS (AMDS), $\hbar$-MDS ($\hbar\geq 2$) and AG    codes to construct new EACQCs. In the previous construction of EACQCs, entanglement-assisted quantum MDS  (EAQMDS) codes were used as the outer codes. Although AMDS and $\hbar$-MDS codes are not optimal compared to MDS codes, they have much larger code length range than MDS codes \cite{qian2019constructions,de1996almost,dodunekov1994near,landjev2015main}. Therefore we use AMDS and $\hbar$-MDS  codes to construct the outer codes in EACQCs since there are much more varieties than by using MDS codes to construct the outer codes. We construct EACQCs with better parameters than the   best known EAQECCs and QECCs of the same length and net transmissions. By the way, classical AG  codes are one of the main linear codes for constructing    AMDS codes and $\hbar$-MDS codes, thus we use AG codes to construct the outer codes in EACQCs. On the other hand, we show that     EACQCs consume maximal entanglement if both the inner and outer codes are with maximal entanglement. Then we construct several new maximal-entanglement EACQCs by concatenating two EAQECCs with maximal entanglement. In particular, we construct three new optimal EACQCs with maximal entanglement. Moreover, we also derive several new maximal-entanglement EACQCs whose minimum distance is only one less than the minimum distance of optimal codes. Furthermore,  according to \cite{carlet2018linear,pereira2021entanglement}, each nonbinary linear code   is equivalent  an LCD code and thus can be used to construct an EAQECC with maximal entanglement.  We use AG codes    exceeding the Gilbert-Varshamov (GV) bound \cite{tsfasman2013algebraic} to construct  the  maximal-entanglement outer codes  in EACQCs. As a result, we explicitly construct two new families of asymptotically good EACQCs with maximal entanglement by using EAQAG codes as the outer codes. At last, we generalize the asymptotic result of Blokh and Zyablov  \cite{blokh1973existence} from classical concatenate codes   to EACQCs. We show that there exist asymptotically good EACQCs meeting the   entanglement-assisted quantum GV bound.



This paper is organized as follows. In Section \ref{Sec2}, we give the   preliminaries and basic knowledge   of classical linear codes and EAQECCs. In Section \ref{Sec3}, we construct new EACQCs with parameters better than the best known quantum codes.  In Section \ref{Sec4EACQCMaximalEntanglement}, we construct new EACQCs with maximal entanglement. In Section \ref{AsymptoticallygoodEACQCs}, we prove that EACQCs can attain the entanglement-assisted quantum GV bound asymptotically. At last, the conclusion and discussion are given in Section \ref{ConclusionandDiscussion}.


\section{Preliminaries} \label{Sec2}

In this section, we first  review some basic knowledge  about entanglement-assisted quantum error-correcting codes and entanglement-assisted concatenated quantum codes. We also introduce the  basic definitions and notations of some specific families of classical  linear codes and   classical algebraic geometry codes.

\subsection{Entanglement-Assisted Quantum Error-Correcting Codes}

Let $p$ be a prime   and let $q$ be a prime power of $p$, i.e., $q = p^{m_0} $ for some positive integer $m_0$. Denote by $\mathbb{F}_q$   the Galois   field  of size $q$   and denote by    $ \mathbb{F}_{q^m} $      the extension  field of $\mathbb{F}_q$, where $m  $ is a positive integer.    Denote  $\mathbb{C}$ by  the field of complex numbers.
Denote  the complex Hilbert space  by $V_n=(\mathbb{C}^q)^{\otimes n}=\mathbb{C}^{q^n}$, which is  the $n$th tensor
product of $\mathbb{C}^q$, where $n$ is a positive integer. Let   $a$ and $b$ be two vectors of  $\mathbb{F}_q^n$.  Let $\xi=\exp(2\pi \textrm{i}/p)$ be a primitive $p$th root of unity. For a quantum state $| \psi\rangle\in V_n$, we define  $X(a)| \psi\rangle=|a+\psi\rangle$ and $Z(b)|\psi\rangle=\xi^{Tr(b\psi)}| \psi\rangle$, where ``$Tr(\cdot)$'' is the trace operation from $\mathbb{F}_q$ to $\mathbb{F}_p$.  Denote by the group
$
\mathcal{G}_n=\{\xi^\imath X(a)Z(b)|  a,b\in\mathbb{F}_q^n, \imath\in \mathbb{F}_q\}
$. For any $\mathcal{E}=\xi^\imath X(a)Z(b)\in \mathcal{G}_n$ with $a=(a_1,\ldots,a_n) $ and $b=(b_1,\ldots,b_n) $, the quantum  weight of $\mathcal{E}$ is defined as \begin{equation}
\textrm{wt}_Q(\mathcal{E})=|\{1\leq i\leq n| (a_i,b_i)\neq(0,0)\}|.
 \end{equation}

  Before the introduction   of EAQECCs, we need to   review some basic definitions and notations of standard QECCs in the language of coding theory.   Denote by  $Q=[[n, k,d]]_q$  a quantum error-correcting code which is a
$q^k$-dimensional subspace of the complex Hilbert space $\mathbb{C}^{q^n}$. The QECC $Q$ has a minimum distance $d$ means that it can detect any quantum error over $\mathbb{C}^{q^n}$ with quantum weight less than $d$.   Let   $| \alpha\rangle  $ and $| \beta\rangle $ be any two quantum states of the quantum code  $Q$, and let $ \mathcal{E} \in \mathcal{G}_n$ be any quantum error such that $1\leq \textrm{wt}_Q(\mathcal{E})\leq d-1$. If $\langle \alpha | \mathcal{E} |\beta \rangle =0 $, then $Q$ is nondegenerate, otherwise it is degenerate.
The standard quantum error-correcting codes can be constructed from classical linear codes satisfying certain dual-containing constraint. Such condition can not be maintained by classical error-correcting codes directly and greatly restricts  the construction of quantum codes.  The entanglement-assisted quantum error-correcting codes can be constructed from any classical linear codes without the dual-containing constraint. From the view of coding theory, EAQECCs can be seen as a further generalization of the standard QECCs by exploiting some extra entanglement.  An EAQECC with parameters $Q_e=[[n_e,k_e,d_e;c]]_q$ is a
$q^{k_e}$-dimensional subspace of $\mathbb{C}^{q^{n_e}}$  by consuming $c$ pre-shared pairs of maximally entangled
states
  between the sender  and the receiver.  The minimum distance of $Q_e$ is
$d_e$ and it can detect any quantum error of weight less than $d_e$.
Since the pre-shared   entanglement  in EAQECCs is extra resource  has to be paid during the transmission, we need to consider the  \emph{net}  transmission of $Q_e$  in a sense, i.e., $k^*=k-c$. In particular, if  $c=n-k$, then $Q_e$ is an EAQECC with maximal entanglement. By generalizing    the construction of standard QECCs, e.g., the
   Calderbank-Shor-Steane (CSS) code     \cite{steane1996error,calderbank1998quantum} and the Hermitian QECC  \cite{calderbank1998quantum}, there exist the following       EAQECCs derived from classical linear codes.

 Denote by $C = [n,k,d]_q$  a classical linear code with length $n$, dimension $k$ and minimum distance $d$ over $\mathbb{F}_q$.  Let $H = (a_{ij})_{(n-k)\times n}$ be the parity-check matrix of $C$.   The dual code of $C$ is defined as $C^\bot = \{u\in \mathbb{F}_q^n|uH^T=0\}$.  Let $\mathbf{C} = [\mathbf{n},\mathbf{k},\mathbf{d}]_{q^2}$  be a linear code over $\mathbb{F}_{q^2}$.  Denote by $\mathbf{H} = (\mathbf{a}_{ij})_{(\mathbf{n}-\mathbf{k})\times \mathbf{n}}$   the parity-check matrix of $\mathbf{C}$.  Denote by the  conjugate transpose
of $\mathbf{H}$ as $\mathbf{H}^{\dagger} = (  \mathbf{a}_{ji}^q)_{ \mathbf{n} \times (\mathbf{n}-\mathbf{k})}$. The Hermitian dual code of $\mathbf{C}$ is defined as $\mathbf{C}^{\bot_h} = \{\mathbf{u}\in \mathbb{F}_{q^2}^n|\mathbf{u}\mathbf{H}^\dagger =0\}$.
  \begin{lemmas}[\cite{brun2006correcting,wilde2008optimal,galindo2019entanglement}]
  \label{EuclidEAQECCs}
     Let $C_1 = [n,k_1,d_1]_q$ and $C_2 = [n,k_2,d_2]_q$ be two classical linear codes over $\mathbb{F}_{q}$. Denote the parity-check matrices of $C_1$ and $C_2$ by $H_1$ and $H_2$, respectively. There exists an EAQECC   with parameters $ {Q}_e=[[n,k_1+k_2-n+c,d_e \geq d ;c]]_q$, where $c=rank(HH^T)=\dim(C_2^{\perp})-\dim(C_2^{\perp} \cap C_1 )$. If $C_2^\bot \subseteq C_1$, then  $ {Q}_e$ is a standard QECC with $c=0$.
 \end{lemmas}
\begin{lemmas}[\cite{brun2006correcting,wilde2008optimal,galindo2019entanglement}]\label{HermitianEAQECCs}
    Let $\mathbf{C}=[\mathbf{n},\mathbf{k},\mathbf{d}]_{q^2}$ be a classical linear code over $\mathbb{F}_{q^2}$. Let the parity-check matrix of $\mathbf{C}$ be $\mathbf{H}$. There exists an EAQECC   with parameters $ \mathbf{Q}_e = [[\mathbf{n},2\mathbf{k}-\mathbf{n}+\mathbf{c},\mathbf{d}_e \geq \mathbf{d};\mathbf{c}]]_q$, where  $\mathbf{c}= rank(\mathbf{H}\mathbf{H}^{\dagger})=\dim(\mathbf{C}^{\perp_h})-\dim(\mathbf{C}^{\perp_h} \cap \mathbf{C} )$. If $\mathbf{C}^{\bot_h} \subseteq \mathbf{C} $, then  $\mathbf{Q}_e$ is a standard QECC with $\mathbf{c}=0$.
\end{lemmas}

Compared to standard QECCs, EAQECCs have to establish some entanglement before the transmission. Although the pre-shared entangled states  can enhance the transmission rate of EAQECCs, they are extra physical resources  cost  from the perspective of quantum information theory.   Therefore we will compare the parameters of EAQECCs and standard QECCs with respect to the net transmission as in  \cite{brun2006correcting,brun2014catalytic}. Recently, the concept of concatenated quantum codes was generalized to entanglement-assisted concatenated quantum codes \cite{fan2022entanglement}. By selecting different inner and outer codes, several families  of EACQCs with parameters better than the best known EAQECCs were constructed.
\begin{lemmas}\cite{fan2022entanglement}\label{EACQC scheme}
    Let $ {Q}_I=[[n_1,k_1,d_1;c_1]]_q$ be the inner EAQECC, and let $ {Q}_O=[[n_2,k_2,d_2;c_2]]_{q^{k_1}}$ be the outer EAQECC. There exists an EACQC $ {Q}_e$ with parameters given by
    \begin{equation}
        \mathcal{Q}_e=[[n_1n_2,k_1k_2,d_e \geq d_1d_2;c_e]]_q,
    \end{equation}
    where $c_e= c_1n_2+c_2k_1$ is the number of pre-shared maximally entangled states. The net transmission is denoted by $k_{e}^{*}=k_1k_2-c_e$.
\end{lemmas}

\subsection{Linear Codes and Algebraic Geometry Codes}
A $q$-ary classical linear code with parameters $C=[n,k,d]_q$ should follow the Singleton bound such that $d\leq n-k+1$. Denote  $ \hbar=n-k+1-d$ by the Singleton defect of $C$. If  $C$ can attain the Singleton bound, i.e., $ \hbar=0$, then $C$ is   called a  maximum-distance-separable (MDS) code. Classical MDS codes are a significant family of  linear codes since they can attain the upper bound and are optimal codes.  If $ \hbar=1$, then $C$ is called an almost MDS   code since it can   almost attain the Singleton bound and is almost optimal. Furthermore, if the dual code of an AMDS code is also an AMDS code, then such AMDS codes are called near-MDS (NMDS) codes. If $ \hbar\geq2$, then we call $C$ as an $\hbar$-MDS code.

Although AMDS codes and $\hbar$-MDS codes are not as good as MDS codes in code parameters, they have a much larger upper bound on code length than MDS codes.    The MDS conjecture \cite{seroussi1986mds} gives that the code length of any $q$-ary MDS code   satisfies $n\leq q+1$  except for the case   $k=3$ and $k=q-1$ with $q$ even and $n=q+2$. The length of a $q$-ary AMDS code could be much larger than $q+2$. Denote by $\mathcal{A}(k,q)$ the maximum length $n$ for which there exists   an $[n,k,n-k]_q$ AMDS code over $\mathbb{F}_q$.  There exist infinite families of AMDS codes constructed  by using  algebraic geometry codes in \cite{tsfasman2013algebraic},
see also \cite{de1996almost,ding2020infinite}.
\begin{lemmas}[{\cite[Theorem 2.3.17]{tsfasman2013algebraic}}]
\label{amdslowerbound}
Let $p$ be a prime and let $q=p^m$, $m$ is a positive integer. For an $[n,k,n-k]_q$ AMDS code and any $2\leq k\leq n-2$, there is
\begin{equation}\mathcal{A}(k,q)\geq
\left\{
             \begin{aligned}
            \chi_q   &\hspace{5mm} \textrm{if}\ p |\lfloor2\sqrt{q}\rfloor\ \textrm{and}\ m\geq3,\ m\ \textrm{odd}\\
            \chi_q +1 & \hspace{5mm}  \textrm{otherwise},
             \end{aligned}
\right.
\end{equation}
where $\chi_q=q+\lfloor2\sqrt{q}\rfloor$.
\end{lemmas}

The bound in Lemma \ref{amdslowerbound} is a famous lower bound for the existence of AMDS  codes. Indeed there exists a more general bound for AMDS codes and $\hbar$-MDS codes in the framework of AG codes. Before introducing it, we first need to  review some basic knowledge of  AG codes.

Let $\mathcal{X} / \mathbb{F}_q$ be an algebraic curve of genus $g$. We define the set of all points
 as $\mathbb{P}_{F} = \{ P|P $ is a point over $\mathcal{X} / \mathbb{F}_q \}$. The divisor group $\textrm{Div}(F)$ is the free abelian group generated by the points over $\mathcal{X}/\mathbb{F}_q$.
In other words, a divisor is a formal sum $D = \textstyle\sum_{P \in \mathbb{P}_F}^{}n_{P}P$, where $n_{P} \in \mathbb{Z}$,  and almost all $n_{P} = 0$. The support and degree of $D$ are defined as $ \textrm{supp}(D) = \{ P \in \mathbb{P}_F | n_{P} \neq 0 \}$ and $\deg(D) =\textstyle\sum_{P \in \mathbb{P}_F}^{} n_{P}  \deg(P)$, respectively, where $\deg(P)$ is the degree of the point. A point of degree one is also called a rational point over $\mathcal{X} / \mathbb{F}_q$. A divisor $G$ over $\mathcal{X}/\mathbb{F}_q$ is said to be effective if $\textrm{supp}(G) \cap \textrm{supp}(D) = \emptyset$.

Denote  $\mathbb{F}_q(\mathcal{X})$ by the function field of $\mathcal{X}$ and denote   the discrete valuation corresponding to  $P$ of $\mathcal{X} / \mathbb{F}_q$ by $\mu_P$.
 For a given divisor $G$, we define the Riemann-Roch space associated to $G$ by $\mathcal{L}(G) = \{ x\in \mathbb{F}_q(\mathcal{X}) \backslash \{0\}|  \textrm{Div}(x)+G\geq  0 \}$.
Denote the dimension  of  $\mathcal{L}(G) $ by $\dim(G)$. According to the Riemann-Roch theorem \cite{tsfasman2013algebraic}, there is $\dim(G)\geq \deg(G)+1-g$.
Let $P_1,\cdots,P_n$ be pairwise distinct rational points of $\mathcal{X} / \mathbb{F}_q$ and denote by $\mathcal{S}=\{P_1,\cdots,P_n\}$. Take a divisor $G$ such that
$\textrm{supp}(G)\cap \mathcal{S}=\emptyset$. Define the map as $\varphi: \mathcal{L}(G)\rightarrow \mathbb{F}_q^n$. The image of $\varphi$ is a subspace of $\mathbb{F}_q^n$ and is denoted by $C_L(G;\mathcal{S})$. If $n$ is larger than the degree of $G$, then $C_L(G;\mathcal{S})$ defines an algebraic geometry code and the dimension of the code is just $\dim(G)$.
\begin{lemmas}[\cite{tsfasman2007algebraic}]
\label{tsfasman2007AGcodes}
$C_L(G;\mathcal{S})$ is an $[n,k,d]_q$ linear code such that $k\geq \deg(G)-g+1$ and $d\geq n-\deg(G)$. If $\deg(G) \geq 2g-1$, then $k=\deg(G)-g+1$.
\end{lemmas}

 Define the asymptotic quantity as follows
\begin{equation}
A(q) = \limsup_{g\rightarrow \infty} \frac{\mathbf{N}_q(g)}{g}
\end{equation}
  where  $\mathbf{N}_q(g)$ is
the maximum number of $\mathbb{F}_q$-rational pionts of a curve $\mathcal{X}/\mathbb{F}_q$ of genus $g$. The asymptotic performance of AG codes is given as follows.
\begin{lemmas}[\cite{tsfasman2007algebraic}]
\label{asymptoticTVZbound}
 There exists a family of AG codes $C=[n,k,d]_q$ with
\begin{equation}
\frac{k}{n}\geq 1-\delta-\frac{1}{A(q)},
\end{equation}
where $0\leq \delta\leq 1$ is the relative minimum distance of $C$. If $q$ is a square, then $A(q)=\sqrt{q}-1$.
\end{lemmas}

The bound in Lemma \ref{asymptoticTVZbound} is called the Tsfasman-Vl\v{a}du\c{t}-Zink (TVZ) bound which is    an essential bound   in  AG codes. The TVZ bound can exceed the classical GV bound when $q\geq49$ is a power of a prime. Moreover, the TVZ bound is also significant in the explicit construction of asymptotically good QECCs \cite{ashikhmin2001asymptotically} and EAQECCs \cite{pereira2021entanglement}.

Denote by $N_q(g) = \max_\mathcal{X}
|\mathcal{X}(\mathbb{F}_q)|$, the maximum being taken over all curves of genus
$g$ over $\mathbb{F}_q$. According to the Weil (or Serre) bound in \cite{tsfasman2007algebraic,tsfasman2013algebraic}, there is
\begin{equation}
N_q(g)\leq q+1+g\lfloor 2\sqrt{q}\rfloor.
\end{equation}
Curves with $N=N_q(g)= q+1+g\lfloor 2\sqrt{q}\rfloor$ are  called maximal. In general, it is difficult to compute the exact value of $N_q(g)$. However, for curves of small genus, e.g., $g=1$ and $g=2$, the exact value of $N_q(g)$ is known. For $g=1$, the value of $N_q(g)$ is given in Lemma \ref{amdslowerbound}. For $g=2$, there is the following result.
\begin{lemmas}[{\cite[Theorem 2.3.18]{tsfasman2013algebraic}}]
\label{hmdslowerbound}
Let $p$ be a prime and let $q=p^m$, $m$ is a positive integer.
\begin{itemize}
\item[1).] If $m$ is even and $q\ne 4,9$, then
$N_q(2)= q+1+   4\sqrt{q} $.
\item[2).] $N_4(2)=10,N_9(2)=20$.
\item[3).] Let $m$ be an odd. Then $q$ is called special iff either $p|\lceil 2\sqrt{q}\rceil$ or $q=m^2+1$, $q=m^2+m+1$, or $q=m^2+m+2$ for some integer $m$. If $q$ is not special, then $N_q(2)= q+1+   2\lceil 2\sqrt{q}\rceil$. If $q$ is   special,
    \begin{equation}N_q(2)=
\left\{
             \begin{aligned}
            q+  2\lceil \sqrt{q}\rceil\ & \textrm{for}\ \varsigma_q>\frac{\sqrt{5}-1}{2},\\
            q+  2\lceil\sqrt{q}\rceil -1\ & \textrm{for}\ \varsigma_q<\frac{\sqrt{5}-1}{2},
             \end{aligned}
\right.
\end{equation}
where $\varsigma_q=2  \sqrt{q} -\lceil2 \sqrt{q}\rceil$.
 \end{itemize}
\end{lemmas}

Different to the cases of $g=1,2$, only some specific values of $q$ are known for $N_q(g)$ with $g=3$, see Theorem 2.3.19 in  \cite{tsfasman2013algebraic}. For example,   $N_{q}(3)$ is equal to $38$ when $g=3$ and  $q=16$.

\begin{table*}[!t]
\caption{New EACQCs with better parameters than the previously best known nondegenerate EAQECCs and standard QECCs. The outer codes are chosen as EAQAMDS  codes.}
\label{table1:EACQCsAMDS}
\centering
\begin{tabular}{llllll}
\toprule
Inner codes & Outer codes & New EACQCs & EAQECCs in \cite{Grassl:codetables} &  QECCs in \cite{Grassl:codetables} \\
$[[n_1,k_1,d_1;c_1]]_2$ & $[[n_2,k_{2}^*,d_2]]_{2^{k_1}}$ & $[[N_1,K_{1}^*,d_e]]_2$ & $[[N_2,K_{2}^*,D_2]]_2$& $[[N_3,K_{3},D_3]]_2$\\
\midrule
$ [[4,2,2;0]]_2$ & $[[23,1^*,11]]_4$ & $[[92,2^*,d_e\geq 22]]_2$  & $[[92,2^*,21]]_2$ & $[[92,2,20]]_2$\\
\multicolumn{2}{c}{Code Extension} & $[[93,2^*,d_e\geq22 ]]_2$ &$[[93,1^*, 21 ]]_2$   & $[[93,1,21]]_2$\\
\multicolumn{2}{c}{Code Extension} & $[[94,2^*,d_e\geq22 ]]_2$ &$[[94,2^*, 21 ]]_2$   & $[[94,1,21]]_2$\\
\multicolumn{2}{c}{Code Expurgation} & $[[91,1^*,d_e\geq22 ]]_2$ &$[[91,1^*, 21 ]]_2$   & $[[91,1,21]]_2$\\
\hline
$ [[4,2,2;0]]_2$ & $[[23,3^*,10]]_4$ & $[[92,6^*,d_e\geq20]]_2$  & $[[92,6^*,19]]_2$ & $[[92,5,19]]_2$\\
\multicolumn{2}{c}{Code Extension} & $[[93,6^*,d_e\geq20 ]]_2$ &$[[93,5^*, 20 ]]_2$   & $[[93,5,19]]_2$\\
\multicolumn{2}{c}{Code Extension} & $[[94,6^*,d_e\geq20 ]]_2$ &$[[94,6^*, 19 ]]_2$   & $[[94,5,19]]_2$\\
\multicolumn{2}{c}{Code Extension} & $[[95,6^*,d_e\geq20 ]]_2$ &$[[95,5^*, 20 ]]_2$   & $[[95,5,19]]_2$\\
\hline
$ [[4,2,2;0]]_2$ & $[[24,2^*,11]]_4$ & $[[96,4^*,d_e\geq22]]_2$  & $[[96,4^*,20]]_2$ & $[[96,4,20]]_2$\\
\multicolumn{2}{c}{Code Extension} & $[[97,4^*,d_e\geq22 ]]_2$ &$[[97,3^*, 20 ]]_2$   & $[[97,3,21]]_2$\\
\multicolumn{2}{c}{Code Extension} & $[[98,4^*,d_e\geq22 ]]_2$ &$[[98,4^*, 20 ]]_2$   & $[[98,3,21]]_2$\\
\multicolumn{2}{c}{Code Expurgation} & $[[95,3^*,d_e\geq22 ]]_2$ &$[[95,3^*, 21 ]]_2$   & $[[95,3,21]]_2$\\
\hline
$ [[4,2,2;0]]_2$ & $[[24,4^*,10]]_4$ & $[[96,8^*,d_e\geq20]]_2$  & $[[96,8^*,19]]_2$ & $[[96,8,19]]_2$\\
\multicolumn{2}{c}{Code Extension} & $[[97,8^*,d_e\geq20 ]]_2$ &$[[97,7^*, 20 ]]_2$   & $[[97,8,19]]_2$\\
 \multicolumn{2}{c}{Code Expurgation} & $[[95,7^*,d_e\geq20 ]]_2$ &$[[95,7^*, 19 ]]_2$   & $[[95,7,19]]_2$\\
 \hline
$ [[4,2,2;0]]_2$ & $[[25,1^*,12]]_4$ & $[[100,2^*,d_e\geq24]]_2$  & $[[100,2^*,22]]_2$ & $[[100,2,21]]_2$\\
  \multicolumn{2}{c}{Code Extension} & $[[101,2^*,d_e\geq24 ]]_2$ &$[[101,1^*, 21 ]]_2$   & $[[101,2,22]]_2$\\
   \multicolumn{2}{c}{Code Extension} & $[[102,2^*,d_e\geq24 ]]_2$ &$[[102,2^*, 21 ]]_2$   & $[[102,2,22]]_2$\\
   \multicolumn{2}{c}{Code Extension} & $[[103,2^*,d_e\geq24 ]]_2$ &$[[103,1^*, 22 ]]_2$   & $[[103,2,22]]_2$\\
   \multicolumn{2}{c}{Code Extension} & $[[104,2^*,d_e\geq24 ]]_2$ &$[[104,2^*, 22 ]]_2$   & $[[104,2,22]]_2$\\
   \multicolumn{2}{c}{Code Extension} & $[[105,2^*,d_e\geq24 ]]_2$ &$[[105,1^*, 23 ]]_2$   & $[[105,2,22]]_2$\\
  \multicolumn{2}{c}{Code Extension} & $[[106,2^*,d_e\geq24 ]]_2$ &$[[106,2^*, 23 ]]_2$   & $[[106,2,22]]_2$\\
 \multicolumn{2}{c}{Code Extension} & $[[107,2^*,d_e\geq24 ]]_2$ &$[[107,1^*, 24 ]]_2$   & $[[107,2,23]]_2$\\
  \multicolumn{2}{c}{Code Expurgation} & $[[99, 1^*,d_e\geq24 ]]_2$ &$[[99,1^*, 21 ]]_2$   & $[[99,1,23]]_2$\\
   \hline
$ [[4,2,2;0]]_2$ & $[[25,3^*,11]]_4$ & $[[100,6^*,d_e\geq22]]_2$  & $[[100,6^*,19]]_2$ & $[[100,6,20]]_2$\\
   \multicolumn{2}{c}{Code Extension} & $[[101,6^*,d_e\geq22 ]]_2$ &$[[101,1^*, 21 ]]_2$   & $[[101,6,19]]_2$\\
   \multicolumn{2}{c}{Code Extension} & $[[102,6^*,d_e\geq22 ]]_2$ &$[[102,2^*, 21 ]]_2$   & $[[102,6,19]]_2$\\
   \multicolumn{2}{c}{Code Extension} & $[[103,6^*,d_e\geq22 ]]_2$ &$[[103,3^*, 21 ]]_2$   & $[[103,6,19]]_2$\\
   \multicolumn{2}{c}{Code Extension} & $[[104,6^*,d_e\geq22 ]]_2$ &$[[104,4^*, 21 ]]_2$   & $[[104,6,20]]_2$\\
   \multicolumn{2}{c}{Code Extension} & $[[105,6^*,d_e\geq22 ]]_2$ &$[[105,5^*, 21 ]]_2$   & $[[105,6,20]]_2$\\
 \multicolumn{2}{c}{Code Extension} & $[[106,6^*,d_e\geq22 ]]_2$ &$[[106,6^*, 21 ]]_2$   & $[[106,6,20]]_2$\\
 \multicolumn{2}{c}{Code Extension} & $[[107,6^*,d_e\geq22 ]]_2$ &$[[107,5^*, 22 ]]_2$   & $[[107,6,21]]_2$\\
\multicolumn{2}{c}{Code Expurgation} & $[[99,5^*,d_e\geq22 ]]_2$ &$[[99,5^*, 20 ]]_2$   & $[[99,5,19]]_2$\\
\bottomrule
\end{tabular}
\end{table*}

\begin{table*}[!t]
\caption{New EACQCs with better parameters than the previously best known nondegenerate EAQECCs and standard QECCs. The outer codes are chosen as $\hbar_e$-EAQMDS  codes with $\hbar_e\leq4$.}
\label{table1:EACQCshbarMDS}
\centering
\begin{tabular}{llllll}
\toprule
Inner codes & Outer codes & New EACQCs & EAQECCs in \cite{Grassl:codetables} &  QECCs in \cite{Grassl:codetables} \\
$[[n_1,k_1,d_1;c_1]]_2$ & $[[n_2,k_{2}^*,d_2]]_{2^{k_1}}$ & $[[N_1,K_{1}^*,d_e]]_2$ & $[[N_2,K_{2}^*,D_2]]_2$& $[[N_3,K_{3},D_3]]_2$\\
\midrule
$ [[4,2,2;0]]_2$ & $[[29,1^*,d_2 \geq13]]_4$ & $[[116,2^*,d_e\geq 26]]_2$  & $[[116,2^*,24]]_2$ & $[[116,2,25]]_2$\\
\multicolumn{2}{c}{Code Extension} & $[[117,2^*,d_e\geq 26]]_2$  & $[[117,1^*,24]]_2$ & $[[117,2,25]]_2$\\
\multicolumn{2}{c}{Code Extension} & $[[118,2^*,d_e\geq26 ]]_2$ &$[[118,2^*, 24 ]]_2$   & $[[118,2,25]]_2$\\
 \multicolumn{2}{c}{Code Expurgation} & $[[115,1^*,d_e\geq 26 ]]_2$ &$[[115,1^*, 24 ]]_2$   & $[[115,1,25]]_2$\\
\hline
$ [[4,2,2;0]]_2$ & $[[29,3^*,10]]_4$ & $[[116,6^*,d_e\geq24]]_2$  & $[[116,6^*,23]]_2$ & $[[116,6,22]]_2$\\
\multicolumn{2}{c}{Code Extension} & $[[117,6^*,d_e\geq24]]_2$  & $[[117,5^*,23]]_2$ & $[[117,6,22]]_2$\\
\multicolumn{2}{c}{Code Extension} & $[[118,6^*,d_e\geq24]]_2$  & $[[118,6^*,23]]_2$ & $[[118,6,22]]_2$\\
\multicolumn{2}{c}{Code Expurgation} & $[[115,5^*,d_e\geq 24 ]]_2$ &$[[115,5^*, 23 ]]_2$   & $[[115,5,23]]_2$\\
\hline
$ [[4,2,2;0]]_2$ & $[[30,2^*,13]]_4$ & $[[120,4^*,d_e\geq26]]_2$  & $[[120,4^*,24]]_2$ & $[[120,4,24]]_2$\\
\multicolumn{2}{c}{Code Extension} & $[[121,4^*,d_e\geq26]]_2$  & $[[121,3^*,25]]_2$ & $[[121,4,24]]_2$\\
\multicolumn{2}{c}{Code Extension} & $[[122,4^*,d_e\geq26]]_2$  & $[[122,4^*,25]]_2$ & $[[122,4,24]]_2$\\
\multicolumn{2}{c}{Code Extension} & $[[123,4^*,d_e\geq26]]_2$  & $[[123,3^*,25]]_2$ & $[[123,4,24]]_2$\\
\multicolumn{2}{c}{Code Expurgation} & $[[119,3^*,d_e\geq26 ]]_2$ &$[[119,3^*, 24 ]]_2$   & $[[119,3,25]]_2$\\
  \hline
$ [[4,2,2;0]]_2$ & $[[30,4^*,12]]_4$ & $[[120,8^*,d_e\geq 24]]_2$  & $[[120,8^*,23]]_2$ & $[[120,8,22]]_2$ \\
\multicolumn{2}{c}{Code Extension} & $[[121,8^*,d_e\geq 24]]_2$  & $[[121,7^*,24]]_2$ & $[[121,8,22]]_2$ \\
\multicolumn{2}{c}{Code Extension} & $[[122,8^*,d_e\geq 24]]_2$  & $[[122,8^*,23]]_2$ & $[[122,8,22]]_2$ \\
\multicolumn{2}{c}{Code Extension} & $[[123,8^*,d_e\geq 24]]_2$  & $[[123,7^*,24]]_2$ & $[[123,8,22]]_2$ \\
 \multicolumn{2}{c}{Code Expurgation} & $[[119,7^*,d_e\geq24 ]]_2$ &$[[119,7^*, 23 ]]_2$   & $[[119,7,22]]_2$ \\
 \hline
$ [[4,2,2;0]]_2$ & $[[31,1^*,14]]_4$ & $[[124,2^*,d_e\geq28]]_2$  & $[[124,2^*,25]]_2$ & $[[124,2,25]]_2$\\
  \multicolumn{2}{c}{Code Extension} & $[[125,2^*,d_e\geq28 ]]_2$ &$[[125,1^*, 26 ]]_2$   & $[[125,2,25]]_2$\\
   \multicolumn{2}{c}{Code Extension} & $[[126,2^*,d_e\geq28 ]]_2$ &$[[126,2^*, 26 ]]_2$   & $[[126,2,25]]_2$\\
   \multicolumn{2}{c}{Code Extension} & $[[127,2^*,d_e\geq28 ]]_2$ &$[[127,1^*, 26 ]]_2$   & $[[127,2,25]]_2$\\
  \multicolumn{2}{c}{Code Expurgation} & $[[123, 1^*,d_e\geq28 ]]_2$ &$[[123,1^*, 26 ]]_2$   & $[[123,1,25]]_2$\\
   \hline
$ [[4,2,2;0]]_2$ & $[[31,3^*,13]]_4$ & $[[124,6^*,d_e\geq26]]_2$  & $[[124,6^*,24]]_2$ & $[[124,6,24]]_2$\\
   \multicolumn{2}{c}{Code Extension} & $[[125,6^*,d_e\geq26]]_2$  & $[[125,5^*,25]]_2$ & $[[125,6,24]]_2$\\
 \multicolumn{2}{c}{Code Expurgation} & $[[123,5^*,d_e\geq26 ]]_2$ &$[[123,5^*, 24 ]]_2$   & $[[123,5,23]]_2$\\
  \hline
$ [[4,2,2;0]]_2$ & $[[31,5^*,12]]_4$ & $[[124,10^*,d_e\geq24]]_2$  & $[[124,10^*,23]]_2$ & $[[124,10,22]]_2$\\
   \multicolumn{2}{c}{Code Extension} & $[[125,10^*,d_e\geq24]]_2$  & $[[125,9^*,24]]_2$ & $[[125,10,22]]_2$\\
   \multicolumn{2}{c}{Code Expurgation} & $[[123,9^*,d_e\geq24 ]]_2$ &$[[123,9^*, 23 ]]_2$   & $[[123,9,22]]_2$\\
   \multicolumn{2}{c}{Code Expurgation} & $[[122,8^*,d_e\geq24 ]]_2$ &$[[122,8^*, 23 ]]_2$   & $[[122,8,22]]_2$\\
   \hline
  $[[4,2,2;0]]_2$ & $[[32,2^*,14]]_4$ & $[[128,4^*,d_e\geq28]]_2$  & $[[128, 4^*, 26]]_2$ & $[[128,4,25]]_2$\\
   \multicolumn{2}{c}{Code Extension} & $[[129,4^*,d_e\geq28]]_2$  & $[[129, 3^*, 26]]_2$ & $[[129,4,25]]_2$\\
   \multicolumn{2}{c}{Code Extension} & $[[130,4^*,d_e\geq28]]_2$  & $[[130, 4^*, 26]]_2$ & $[[130,4,25]]_2$\\
   \multicolumn{2}{c}{Code Expurgation} & $[[127,3^*,d_e\geq28]]_2$ &$[[127, 3^*, 26]]_2$ & $[[127,3,25]]_2$\\
\multicolumn{2}{c}{Code Expurgation} & $[[126,2^*,d_e\geq28]]_2$ &$[[126, 2^*, 26]]_2$ & $[[126,2,25]]_2$\\
\multicolumn{2}{c}{Code Expurgation} & $[[125,1^*,d_e\geq28]]_2$ &$[[125, 1^*, 26]]_2$ & $[[125,1,27]]_2$\\
\hline
 $[[4,2,2;0]]_2$ & $[[32,4^*,13]]_4$ & $[[128,8^*,d_e\geq26]]_2$  & $[[128,8^*,25]]_2$ & $[[128,8,23]]_2$\\
   \multicolumn{2}{c}{Code Extension} & $[[129,8^*,d_e\geq26]]_2$  & $[[129,7^*,25]]_2$ & $[[129,8,24]]_2$\\
   \multicolumn{2}{c}{Code Extension} & $[[130,8^*,d_e\geq26]]_2$  & $[[130,8^*,25]]_2$ & $[[130,8,24]]_2$\\
     \multicolumn{2}{c}{Code Expurgation} & $[[127, 7^*, d_e\geq 26]]_2$ &$[[127, 7^*, 25]]_2$ & $[[126,7,23]]_2$\\
   \multicolumn{2}{c}{Code Expurgation} & $[[126, 6^*, d_e\geq 26]]_2$ &$[[126, 6^*, 25]]_2$ & $[[125,6,23]]_2$\\
   \multicolumn{2}{c}{Code Expurgation} & $[[125, 5^*, d_e\geq 26]]_2$ &$[[125, 5^*, 25]]_2$ & $[[125,5,24]]_2$\\
   \hline
   $ [[4,2,2;0]]_2$ & $[[33,1^*,15]]_4$ & $[[132, 2^*,d_e\geq 30]]_2$  & $[[132,2^*,27]]_2$ & $[[132,2,26]]_2$\\
   \multicolumn{2}{c}{Code Extension} & $[[133,2^*,d_e\geq 30 ]]_2$ &$[[133,1^*, 27 ]]_2$   & $[[133,2, 26  ]]_2$\\
   \multicolumn{2}{c}{Code Extension} & $[[134,2^*,d_e\geq 30 ]]_2$ &$[[134,2^*, 27 ]]_2$   & $[[134,2, 26 ]]_2$\\
      \multicolumn{2}{c}{Code Extension} & $[[135,2^*,d_e\geq 30 ]]_2$ &$[[135,1^*, 28 ]]_2$   & $[[135,2, 27  ]]_2$\\
      \multicolumn{2}{c}{Code Extension} & $[[136,2^*,d_e\geq 30 ]]_2$ &$[[136,2^*, 28 ]]_2$   & $[[136,2, 27  ]]_2$\\
      \multicolumn{2}{c}{Code Extension} & $[[137,2^*,d_e\geq 30 ]]_2$ &$[[137,1^*, 28 ]]_2$   & $[[137,2, 27  ]]_2$\\
      \multicolumn{2}{c}{Code Extension} & $[[138,2^*,d_e\geq 30 ]]_2$ &$[[138,2^*, 28 ]]_2$   & $[[138,2, 28  ]]_2$\\
      \multicolumn{2}{c}{Code Extension} & $[[139,2^*,d_e\geq 30 ]]_2$ &$[[139,1^*, 29 ]]_2$   & $[[139,2, 29  ]]_2$\\
  \multicolumn{2}{c}{Code Expurgation} & $[[131, 1^*, d_e\geq 30]]_2$ &$[[131, 1^*, 27]]_2$ & $[[131, 1, 27]]_2$\\
    \hline
$ [[4,2,2;0]]_2$ & $[[33,3^*,14]]_4$ & $[[132,6^*,d_e\geq28]]_2$  & $[[132,6^*,26]]_2$ & $[[132, 6, 25]]_2$\\
   \multicolumn{2}{c}{Code Extension} & $[[133,6^*,d_e\geq28 ]]_2$ &$[[133,5^*, 26 ]]_2$  & $[[133, 6, 25]]_2$\\
   \multicolumn{2}{c}{Code Extension} & $[[134,6^*,d_e\geq28 ]]_2$ &$[[134,6^*, 26 ]]_2$  & $[[134, 6, 25]]_2$\\
   \multicolumn{2}{c}{Code Extension} & $[[135,6^*,d_e\geq28 ]]_2$ &$[[135,5^*, 26 ]]_2$  & $[[135, 6, 25]]_2$\\
   \multicolumn{2}{c}{Code Extension} & $[[136,6^*,d_e\geq28 ]]_2$ &$[[136,6^*, 27 ]]_2$  & $[[136, 6, 26]]_2$\\
   \multicolumn{2}{c}{Code Extension} & $[[137,6^*,d_e\geq28 ]]_2$ &$[[137,5^*, 27 ]]_2$  & $[[137, 6, 26]]_2$\\
   \multicolumn{2}{c}{Code Extension} & $[[138,6^*,d_e\geq28 ]]_2$ &$[[138,6^*, 27 ]]_2$  & $[[138, 6, 26]]_2$\\
   \multicolumn{2}{c}{Code Extension} & $[[139,6^*,d_e\geq28 ]]_2$ &$[[139,5^*, 27 ]]_2$  & $[[139, 6, 26]]_2$\\
   \multicolumn{2}{c}{Code Extension} & $[[140,6^*,d_e\geq28 ]]_2$ &$[[140,6^*, 27 ]]_2$  & $[[140, 6, 26]]_2$\\
   \multicolumn{2}{c}{Code Extension} & $[[141,6^*,d_e\geq28 ]]_2$ &$[[141,5^*, 28 ]]_2$  & $[[141, 6, 26]]_2$\\
   \multicolumn{2}{c}{Code Expurgation} &$[[131, 5^*, d_e\geq 28]]_2$ &$[[131, 5^*, 26]]_2$ & $[[131, 5, 25]]_2$\\
   \hline
$ [[4,2,2;0]]_2$ & $[[33,5^*,13]]_4$ & $[[132,10^*,d_e\geq 26]]_2$  & $[[132, 10^*, 25]]_2$ & $[[132, 10, 24]]_2$\\
 \multicolumn{2}{c}{Code Extension} & $[[133,10^*,d_e\geq 26 ]]_2$ &$[[133, 9^*, 25 ]]_2$ & $[[133, 10, 24]]_2$\\
  \multicolumn{2}{c}{Code Extension} & $[[134,10^*,d_e\geq 26 ]]_2$ &$[[134, 10^*, 25 ]]_2$ & $[[134, 10, 24]]_2$\\
 \multicolumn{2}{c}{Code Extension} & $[[135,10^*,d_e\geq 26 ]]_2$ &$[[135, 9^*, 26 ]]_2$ & $[[135, 10, 24]]_2$\\
 \multicolumn{2}{c}{Code Expurgation} &$[[131, 9^*, d_e\geq 26]]_2$ &$[[131, 9^*, 25]]_2$ & $[[131, 9, 24]]_2$\\
  \multicolumn{2}{c}{Code Expurgation} &$[[130, 8^*, d_e\geq 26]]_2$ &$[[130, 8^*, 25]]_2$ & $[[130, 8, 24]]_2$\\
\bottomrule
\end{tabular}
\end{table*}

\section{New Constructions of EACQCs with EAQAG Codes As the Outer Codes}\label{Sec3}
\begin{table*}[!t]
\caption{New EACQCs with better parameters than the previously best known nondegenerate EAQECCs and standard QECCs. The outer codes are chosen as $\hbar_e$-EAQMDS  codes with $\hbar_e\leq6$.}
\label{table1:EACQCs3hbarMDS}
\centering
\begin{tabular}{llllll}
\toprule
Inner codes & Outer codes & New EACQCs & EAQECCs in \cite{Grassl:codetables} &  QECCs in \cite{Grassl:codetables} \\
$[[n_1,k_1,d_1;c_1]]_2$ & $[[n_2,k_{2}^*,d_2]]_{2^{k_1}}$ & $[[N_1,K_{1}^*,d_e]]_2$ & $[[N_2,K_{2}^*,D_2]]_2$& $[[N_3,K_{3},D_3]]_2$\\
\midrule

$ [[4,2,2;0]]_2$ & $[[36,2^*,15]]_4$ & $[[144, 4^*, d_e\geq 30]]_2$  & $[[144, 4^*, 29]]_2$ & $[[144, 4, 28]]_2$\\
\multicolumn{2}{c}{Code Extension} & $[[145, 4^*, d_e\geq 30]]_2$  & $[[145, 3^*, 29]]_2$ & $[[145, 4, 28]]_2$\\
\multicolumn{2}{c}{Code Extension} & $[[146, 4^*, d_e\geq 30]]_2$  & $[[146, 4^*, 29]]_2$ & $[[146, 4, 28]]_2$\\
\multicolumn{2}{c}{Code Extension} & $[[147, 4^*, d_e\geq 30]]_2$  & $[[147, 3^*, 30]]_2$ & $[[147, 4, 28]]_2$\\
 \multicolumn{2}{c}{Code Expurgation} & $[[143, 3^*,d_e\geq 30 ]]_2$ &$[[143, 3^*, 29 ]]_2$   & $[[143, 3, 28]]_2$\\
  \hline
$ [[4, 2, 2;0]]_2$ & $[[37, 1^*, 16]]_4$ & $[[148, 2^*, d_e\geq 32]]_2$  & $[[148, 2^*, 30]]_2$ & $[[148, 2, 30]]_2$ \\
\multicolumn{2}{c}{Code Extension} & $[[149, 2^*,d_e\geq 32]]_2$  & $[[149, 1^*, 30]]_2$ & $[[149, 2, 30]]_2$ \\
\multicolumn{2}{c}{Code Extension} & $[[150, 2^*,d_e\geq 32]]_2$  & $[[150, 2^*, 30]]_2$ & $[[150, 2, 30]]_2$ \\
       \hline
$ [[4,2,2;0]]_2$ & $[[37, 3^*, 15]]_4$ & $[[148, 6^*, d_e\geq 30]]_2$  & $[[148, 6^*, 29]]_2$ & $[[148, 6, 28]]_2$\\
  \multicolumn{2}{c}{Code Extension} & $[[149, 6^*,d_e\geq 30]]_2$  & $[[149, 5^*, 29]]_2$ & $[[149, 6, 28]]_2$ \\
\multicolumn{2}{c}{Code Extension} & $[[150, 6^*,d_e\geq 30]]_2$  & $[[150, 6^*, 29]]_2$ & $[[150, 6, 28]]_2$ \\
  \multicolumn{2}{c}{Code Expurgation} & $[[147, 5^*,d_e\geq 30 ]]_2$ &$[[147, 5^*, 29 ]]_2$   & $[[147, 5, 28]]_2$\\
  \multicolumn{2}{c}{Code Expurgation} & $[[146, 4^*,d_e\geq 30 ]]_2$ &$[[146, 4^*, 29 ]]_2$   & $[[146, 4, 28]]_2$\\
 \multicolumn{2}{c}{Code Expurgation} & $[[145, 3^*,d_e\geq 30 ]]_2$ &$[[145, 3^*, 29 ]]_2$   & $[[145, 3, 28]]_2$\\
   \hline
$ [[4, 2, 2; 0]]_2$ & $[[ 38, 2^*, 16 ]]_4$ & $[[152, 4^*, d_e\geq 32]]_2$  & $[[152, 4^*, 30]]_2$ & $[[152, 4, 29]]_2$\\
   \multicolumn{2}{c}{Code Extension} & $[[153, 4^*, d_e\geq 32]]_2$  & $[[153, 3^*, 31]]_2$ & $[[153, 4, 29]]_2$\\
 \multicolumn{2}{c}{Code Extension} & $[[154, 4^*, d_e\geq 32]]_2$ & $[[154, 4^*, 31 ]]_2$   & $[[154, 4, 30]]_2$\\
  \multicolumn{2}{c}{Code Extension} & $[[155, 4^*, d_e\geq 32]]_2$ & $[[155, 3^*, 31 ]]_2$   & $[[155, 4, 30]]_2$\\
 \multicolumn{2}{c}{Code Extension} & $[[156, 4^*, d_e\geq 32]]_2$ & $[[156, 4^*, 31 ]]_2$   & $[[156, 4, 30]]_2$\\
 \multicolumn{2}{c}{Code Extension} & $[[157, 4^*, d_e\geq 32]]_2$ & $[[157, 3^*, 31 ]]_2$   & $[[157, 4, 30]]_2$\\
 \multicolumn{2}{c}{Code Extension} & $[[158, 4^*, d_e\geq 32]]_2$ & $[[158, 4^*, 31 ]]_2$   & $[[158, 4, 30]]_2$\\
 \multicolumn{2}{c}{Code Extension} & $[[159, 4^*, d_e\geq 32]]_2$ & $[[159, 3^*, 31 ]]_2$   & $[[159, 4, 30]]_2$\\
 \multicolumn{2}{c}{Code Expurgation} & $[[151, 3^*,d_e\geq 32 ]]_2$ &$[[151, 3^*, 30 ]]_2$   & $[[151, 3, 29]]_2$\\
   \hline
$ [[4,2,2;0]]_2$ & $[[38, 4^*, 15]]_4$ & $[[152, 8^*, d_e\geq 30]]_2$  & $[[152, 8^*, 29]]_2$ & $[[152, 8, 28]]_2$\\
   \multicolumn{2}{c}{Code Extension} & $[[153, 8^*, d_e\geq 30]]_2$  & $[[153, 7^*, 30]]_2$ & $[[153, 8, 28]]_2$\\
   \multicolumn{2}{c}{Code Expurgation} & $[[151, 7^*, d_e\geq 30]]_2$ &$[[151, 7^*, 29 ]]_2$   & $[[151, 7, 28]]_2$\\
   \bottomrule
\end{tabular}
\end{table*}
Recall that EAQECCs can be constructed from any classical linear code directly without the restriction of dual-containing relationships. In general, the exact number of preshared entangled quantum states is unknown and  some explicit computations are need. However such computation can be done with the Gaussian elimination  method in polynomial time complexities for classical linear codes \cite{wilde2008optimal,brun2006correcting}. In addition, it was shown in \cite{carlet2018linear} that any   linear code over $\mathbb{F}_q$ ($q>3$) is
equivalent to an Euclidean LCD code and  any linear code over $\mathbb{F}_{q^2}$ ($q>2$)  is equivalent to a Hermitian LCD code. Therefore any linear code  over $\mathbb{F}_q$ ($q>3$)
leads to an Euclidean type $q$-ary EAQECC with maximal entanglement. Also, any linear code over $\mathbb{F}_{q^2}$ ($q>2$) leads to  a Hermitian type $q$-ary EAQECC with maximal entanglement. In such cases, the number of preshared entangled quantum states can be derived determinately.

For a $Q_e=[[n_e,k_e,d_e;c]]_q$ EAQECC, we denote
\begin{equation}
\hbar_e=n_e-k_e-2d_e+2+c
 \end{equation}
 by the EA quantum Singleton defect of $Q_e$. If $\hbar_e=0$, then $Q_e$ is an EAQMDS code. If $\hbar_e=2$, we call  $Q_e $   as an entanglement-assisted quantum AMDS (EAQAMDS) code and then  $d_e=\lfloor(n_e+c-k_e)/2\rfloor$. If $\hbar_e>2$, then we call   $Q_e $   as   an   $\hbar_e$-EAQMDS    code.
 Notice that an $\hbar$-MDS code can lead to an $\hbar_e=2\hbar$-EAQMDS code  by using the CSS or Hermitian constructions.

 Denote by $\mathcal{B}(k_e,q)$ the maximum length $n_e$ for which there exists   an $[[n_e,k_e,d_e=\lfloor(n_e+c-k_e)/2\rfloor;c]]_q$ EAQAMDS code over $\mathbb{F}_q$. We have the following general result  for the lower bound of the maximum length of EAQAMDS codes.
\begin{lemmas}
\label{generalEAQAMDS}
Let $p$ be a prime and let $q=p^m$, $m$ is a positive integer. There exist infinite families of $[[n_e,k_e,d_e=\lfloor(n_e+c-k_e)/2\rfloor;c]]_q$ EAQAMDS codes with $0\leq c\leq n_e-k_e$ such that for any $1\leq k_e\leq n_e-4$, we have
\begin{equation}\mathcal{B}(k_e,q)\geq
            q^2+2q +1.
\end{equation}
\end{lemmas}
\begin{IEEEproof}
Suppose that there exists a $C=[n_e,k,n_e-k]_{q^2}$ classical  AMDS code over $\mathbb{F}_{q^2}$.   Denote  the parity-check matrix of $C$ by $H$.
            There exists a $Q_e=[[n_e,k_e,d_e;c]]_{q}$ EAQECC  according to Lemma \ref{HermitianEAQECCs}, where $k_e=2k-n_e+c$, $d_e=\lfloor(n_e+c-k_e)/2\rfloor$ and $c=rank(HH^\dagger)$. Then we have $\mathcal{B}(k_e,q )\geq\mathcal{A}(k,q^2)\geq
            q^2+2q +1  $ according to Lemma \ref{amdslowerbound}.
\end{IEEEproof}

If $m\geq2$, then there exists infinite families of EAQAMDS codes with maximal entanglement according to \cite{carlet2018linear}.
\begin{corollarys}
Let $p$ be a prime and let $q=p^m$, $m\geq2$ is an   integer. There exists infinite families
of $[[n_e,k_e, n_e -k_e;n_e -k_e]]_q$ EAQAMDS codes with maximal entanglement such that  $\mathcal{B}(k_e,q)\geq
            q^2+2q +1$ for any $1\leq k_e\leq n_e-4$.
\end{corollarys}

 Denote by $\mathcal{C}(k_e,q)$ the maximum length $n_e$ for which there exists   an   $\hbar_e$-EAQMDS code with parameters $[[n_e,k_e,d_e ;c]]_q$ over $\mathbb{F}_q$. We have the following   result  for the lower bound of the maximum length of $\hbar_e$-EAQMDS codes.
\begin{lemmas}
\label{generalEAQAMDS}
Let $p$ be a prime and let $q=p^m$, $m$ is a positive integer. There exist infinite families of $\hbar_e$-EAQMDS codes with parameters $[[n_e,k_e,d_e ;c]]_q$, where $\hbar_e\leq4$ and $0\leq c\leq n_e-k_e$. Moreover,   if $q\ne 2,3$, then  we have
$\mathcal{C}(k_e,q)\geq q^2+4q +1$. If $q =2$, then $\mathcal{C}(k_e,q)\geq10$. If $q = 3$, then $\mathcal{C}(k_e,q)\geq 20$.
\end{lemmas}
\begin{IEEEproof}
According to Lemma \ref{tsfasman2007AGcodes}, there exists a $C=[n,k,d]_{q^2}$-AG codes such that $k\geq\deg(G)-g+1$ and $d\geq n-\deg(G)$. If $\deg(G)\geq 2g-1$, then $k=\deg(G)-g+1$.
Let $g=2$, then the maximum length $n$ of $C$ is $N_{q^2}(2)$. Therefore if $\deg(G)\geq 3$, then we have $C=[n,\deg(G)-1,d\geq n-\deg(G)]_{q^2}$. This code is at least a  $2$-MDS code.
 From Lemma \ref{hmdslowerbound}, we know that
$N_{q^2}(2)= q^2+   4q+1  $ if  $q\ne 2,3$. If $q=2$, then $N_{4}(2)= 10  $ and if $q=3$, then $N_{9}(2)= 20  $. Denote  the parity-check matrix of $C$ by $H$.
            There exists a $Q_e=[[n_e,k_e,d_e;c]]_{q}$ EAQECC  according to Lemma \ref{HermitianEAQECCs}, where $k_e=2k-n_e+c$, $d_e = (n_e-k_e-\hbar_e+c)/2+1$ with $\hbar_e\leq 4$ and $c=rank(HH^\dagger)$. The maximum length of $Q_e$ is given by $N_{q^2}(2)$ from Lemma \ref{hmdslowerbound}.
\end{IEEEproof}

If $m\geq2$, then there exists infinite families of $\hbar_e$-EAQMDS codes with maximal entanglement according to \cite{carlet2018linear}.
\begin{corollarys}
Let $p$ be a prime and let $q=p^m$, $m\geq2$ is an   integer. There exists infinite families
of maximal-entanglement $[[n_e,k_e, d_e;n_e -k_e]]_q$  $\hbar_e$-EAQMDS codes with $\hbar_e\leq4$. If $q\ne 2,3$, then  we have
$\mathcal{C}(k_e,q)\geq q^2+4q +1$. If $q =2$, then $\mathcal{C}(k_e,q)\geq10$. If $q = 3$, then $\mathcal{C}(k_e,q)\geq 20$.
\end{corollarys}

Compared with EAQMDS codes, EAQAMDS codes and $\hbar_e$-EAQMDS codes have a much larger code length with  a EA quantum Singleton defect of   $2$ or more. Therefore we can use EAQAMDS   and $\hbar_e$-EAQMDS codes as the outer   codes to construct   EACQCs with a much more freedom on code length than by using EAQMDS codes. We give an explicit example to illustrate the construction. We chose a binary $Q_1=[[4,2,2]]_2$ QECC as the inner code. Thus the field size of the outer code is ${2^2}$. Then there exists an $[[n_e,k_e,d_e=\lfloor(n_e+c-k_e)/2\rfloor;c]]_q$ EAQAMDS code with length $n_e$ up to $25$ for any $1\leq k_e\leq 21$. We chose  $Q_2=[[25,1+c,12;c]]_4$ as the outer component code. At last we can derive a binary   $Q_e=[[100,2+2c,24;2c]]_2$ EACQC where $0\leq  c\leq 12$. The net transmission of $Q_e$ is $k_e^*=2$. According to the online code tables in \cite{Grassl:codetables}, there exists a best known nondegenerate EAQECC with parameters $[[100,37,21;35]]_2$ and exists a best known $[[100,2,22]]_2$ QECC.  Therefore our new EACQC  has a larger minimum distance than the best known nondegenerate EAQECCs and QECCs of the same length and net transmissions.  In Table \ref{table1:EACQCsAMDS}, we list new  EACQCs constructed by using the $[[4,2,2]]_2$ QECC as the inner code and using  EAQAMDS codes as the outer codes. Moreover, we use the code extension to derive more EACQCs with better parameters than the previously known results. If we use the $[[3,2,2;1]]_2$ EAQECC to replace one of the inner codes, we can also get new EACQCs with shorter code length and net transmissions but maintaining the minimum distance. We call such method as the code expurgation.

In Table \ref{table1:EACQCshbarMDS} and Table \ref{table1:EACQCs3hbarMDS}, we list new  EACQCs constructed by using the $[[4,2,2]]_2$ QECC as the inner code and using  $\hbar_e$-EAQMDS codes  as the outer codes. The EA quantum Singleton defects  of $\hbar_e$-EAQMDS codes in Table \ref{table1:EACQCshbarMDS} and  in Table \ref{table1:EACQCs3hbarMDS} satisfy $\hbar_e\leq 4$ and $\hbar_e\leq 6$, respectively.

\section{EACQCs With Maximal Entanglement }\label{Sec4EACQCMaximalEntanglement}
Maximal-entanglement EAQECCs   can achieve the EA quantum capacity of quantum channels and are also highly related to   classical LCD codes. In this section, we exhibit the performance of maximal-entanglement EACQCs. Firstly, we show that the resultant EACQC  is   maximal-entanglement if both the inner code and the outer code are maximal-entanglement EAQECCs.

\begin{lemmas}\label{Maximal-EntanglementEACQC}
    Let the inner and the outer codes  be two maximal-entanglement EAQECCs   with parameters $Q_1=[[n_1,k_1,d_1;n_1-k_1]]_q$ and $Q_2=[[n_2,k_2,d_2;n_2-k_2]]_{q^{k_1}}$, respectively.   There exists a maximal-entanglement EACQC with parameters
    \begin{equation} {Q}_e=[[n_1n_2,k_1k_2,d_e\geq d_1d_2;n_1n_2-k_1k_2]]_q.
    \end{equation}
\end{lemmas}

\begin{IEEEproof}
According to Lemma \ref{EACQC scheme}, we can obtain an EACQC with parameters $ {Q}_e=[[n_1n_2,k_1k_2,d_e\geq d_1d_2;c_e]]_q$. The number of maximally entangled states is given as $c_e=(n_1-k_1)n_2+(n_2-k_2)k_1=n_1n_2-k_1k_2$. Therefore the resultant EACQC is a maximal-entanglement EAQECC.
\end{IEEEproof}

Let $n_1\geq3$ and $n_2\geq3$  be arbitrary two odd integers. Then there exist two maximal-entanglement EAQECCs with parameters $Q_1=[[n_1,1,n_1;n_1-1]]_2$ and $Q_2=[[n_2,1,n_2;n_2-1]]_2$, respectively \cite{lai2017linear}. By Lemma \ref{Maximal-EntanglementEACQC}, there exists an optimal  $Q_e=[[n_1n_2,1,n_1n_2;n_1n_2-1]]_2$ EACQC with maximal entanglement. This code $Q_e$ belongs  to the family of repetition maximal-entanglement EAQECCs with odd length  \cite{lai2017linear}.
\begin{corollarys}
   There exists a family of  maximal-entanglement EACQCs with  parameters  $Q_e=[[n_1n_2, 1, n_1n_2;n_1n_2-1]]_2$ that are optimal, where $n_1\geq3$ and $n_2\geq3$  are two odd integers.
\end{corollarys}

If we chose other optimal EAQECCs with maximal entanglement in \cite{lai2017linear} as the component codes in EACQCs, we can derive    EACQCs with parameters achieving or close to the upper bound of linear codes in \cite{Grassl:codetables}. For example, we chose $Q_1=[[3,2,2;1]]_2$ as the inner code and chose $Q_2=[[2,1,2;1]]_4$  as the outer code. Then we can derive a maximal-entanglement  EACQC with parameters $Q_e=[[6,2,4;4]]_2$ which is   optimal   according to \cite{lai2017linear}. In Table \ref{table2: Maximal entanglement EACQC}, we construct  more maximal-entanglement EACQCs which are optimal or almost optimal.  In particular, we   derive  three new maximal-entanglement EACQCs with parameters   $[[26,2,20;24]]_2$, $[[36,2,28;34]]_2$, and $[[46,2,36;34]]_2$ that can achieve the upper bound in \cite{Grassl:codetables} and are thus optimal.

\begin{table}[!t]
\caption{Maximal-entanglement EACQCs that are optimal or almost optimal. The bold numbers in the brackets are the minimum distance of the corresponding optimal classical quaternary codes in  \cite{Grassl:codetables}. We set $\Delta_1=n_1-k_1$, $\Delta_2=n_2-k_2$ and $\Delta_e=n_e-k_e$ for the inner codes, the outer codes, and the EACQCs, respectively.}
\label{table2: Maximal entanglement EACQC}
\centering
\begin{tabular}{lll}
\toprule
Inner codes & Outer codes &   EACQCs    \\
$[[n_1,k_1,d_1;\Delta_1]]_2$ & $[[n_2,k_{2},d_2;\Delta_2]]_{2^{k_1}}$ & $[[n_e,k_e,d_e;\Delta_e]]_2$    \\
\midrule
$[[3,2,2;1]]_2$  & $[[2,1,2;1]]_4$ & $[[6,2,\textbf{4}(\textbf{4});4]]_2$    \\
$[[7,2,5;5]]_2$  & $[[2,1,2;1]]_4$ & $[[14,2,\textbf{10}(\textbf{10});12]]_2$    \\
$[[8,2,6;6]]_2$  & $[[2,1,2;1]]_4$ & $[[16,2,\textbf{12}(\textbf{12});14]]_2$    \\
$[[13,2,10;11]]_2$  & $[[2,1,2;1]]_4$ & $[[26,2,\textbf{20}(\textbf{20});24]]_2$    \\
$[[18,2,14;16]]_2$  & $[[2,1,2;1]]_4$ & $[[36,2,\textbf{28}(\textbf{28});34]]_2$   \\
$[[23,2,18;21]]_2$  & $[[2,1,2;1]]_4$ & $[[46,2,\textbf{36}(\textbf{36});34]]_2$    \\
$[[18,2,14;16]]_2$  & $[[3,1,3;2]]_4$ & $[[54,2,42(\textbf{43});52]]_2$   \\
$[[23,2,18;21]]_2$  & $[[3,1,3;2]]_4$ & $[[69,2, 54 (\textbf{55});67]]_2$   \\
$[[23,2,18;21]]_2$  & $[[4,1,4;3]]_4$ & $[[92,2, 72 (\textbf{73});90]]_2$    \\
$[[13,2,10;11]]_2$  & $[[3,1,3;2]]_4$ & $[[39,2, {30}(\textbf{31});37]]_2$   \\
$[[8,2,6;6]]_2$  & $[[3,1,3;2]]_4$ & $[[24,2, {18}(\textbf{19});22]]_2$      \\
$[[8,2,6;6]]_2$  & $[[4,1,4;3]]_4$ & $[[32,2, {24}(\textbf{25});30]]_2$    \\
$[[7,2,5;5]]_2$  & $[[3,1,3;2]]_4$ & $[[21,2, {15}(\textbf{16});19]]_2$    \\
$[[18,2,14;16]]_2$  & $[[4,1,4;3]]_4$ & $[[72,2,56(\textbf{57});70]]_2$    \\

\bottomrule
\end{tabular}
\end{table}

Next, we construct asymptotically  good binary EACQCs with maximal entanglement by using AG codes. According to Lemma \ref{Maximal-EntanglementEACQC}, we can construct EACQCs with maximal entanglement by using the inner and outer codes both with maximal entanglement. In our construction, we first let the outer code of EACQCs be EAQAG codes with maximal entanglement. Then we let the inner codes be some certain EAQECCs to construct EACQCs with maximal entanglement. According to \cite{carlet2018linear,pereira2021entanglement}, there exist explicit constructions of $q$-ary ($q>3$) asymptotically good EAQAG codes with maximal entanglement.
\begin{lemmas}[\cite{carlet2018linear,pereira2021entanglement}]
\label{asymptoticmaximalentanglementEAQAG}
Let $q \geq 4$ be a power of a prime. There exists a family
of asymptotically good EAQAG  codes consuming maximal
entanglement with parameters $\mathcal{Q}=[[n, k, d; n-k]]_q$, such that
\begin{equation}\label{TVZBound}
R=\frac{k}{n}\geq 1-\delta-\frac{1}{A(q)},
\end{equation}
where $0\leq    \delta \leq 1 -  1/ A(q)  $ is  the relative minimum distance.
\end{lemmas}

 We take EAQAG codes with maximal entanglement in Lemma \ref{asymptoticmaximalentanglementEAQAG} as the outer codes and      binary EAQECCs with maximal entanglement  as the inner codes to construct EACQCs. In particular,  we chose $Q_1=[[m_1,m_1,1;0]]_2$  ($m_1>1$ is an integer) and chose $Q_2=[[m_2,m_2-1,2;1]]_2$ ($m_2>2$ is odd) \cite{lai2017linear,brun2014catalytic} as the inner codes in the  construction. Then we can derive EACQCs with maximal entanglement according to Lemma \ref{Maximal-EntanglementEACQC} since $Q_1$ and $Q_2$ are both binary EAQECCs with maximal entanglement.
\begin{propositions}
\label{propositionmaxEACQC}
There exist two families of  EACQCs $\mathcal{Q}_1$ and $\mathcal{Q}_2$ consuming maximal entanglement   with parameters
\begin{equation}
 \mathcal{Q}_1  = [[m_1N_1,m_1K_1,D_1;m_1(N_1-K_1)]]_2,\    \textrm{and}
\end{equation}
\begin{equation}
\mathcal{Q}_2 =[[m_2N_2,(m_2-1)K_2,2D_2;m_2(N_2-K_2)+K_2]]_2,
\end{equation}
  respectively, where $m_1>1$ is an integer and $m_2>2$ is odd. The asymptotic parameters of $\mathcal{Q}_1$ and $\mathcal{Q}_2$ are given as
\begin{eqnarray}
\label{asymptoticGV}
 R_{\mathcal{Q}_1}  &=&\frac{K_1}{N_1} \geq  1-m_1\delta_1-\frac{1}{A(2^{m_1})},\    \textrm{and}  \\
 R_{\mathcal{Q}_2}  &=&\frac{(m_2-1)K_2}{m_2N_2}\\
  &\geq&  \left(1-\frac{1}{m_2}\right)\left(1-\frac{m_2}{2}\delta_1-\frac{1}{A(2^{(m_2-1)})}\right),
 \end{eqnarray}
respectively, where $0\leq\delta_1\leq(1 -1/A(2^{m_1}))/m_1$ and $0\leq\delta_2\leq 2(1 -1/A(2^{(m_2-1)}))/m_2$ are the relative minimum distances of $\mathcal{Q}_1$  and $\mathcal{Q}_2$, respectively.
\end{propositions}
\begin{IEEEproof}
We take the inner EAQECC as $Q_I=[[m_1,m_1,1;0]]_2$ and then take the outer EAQECC as the EAQAG code with parameters $Q_O=[[N_1, K_1,D_1;N_1-K_1]]_{2^{m_1}}$ according to Lemma \ref{Maximal-EntanglementEACQC}. Then we can derive an EACQC with parameters
$\mathcal{Q}_1  = [[m_1N_1,m_1K_1,D_1;m_1(N_1-K_1)]]_2$ according to Theorem \ref{EACQC scheme}. Then  we can derive the asymptotic bound of $
 \mathcal{Q}_1$ from Lemma \ref{asymptoticmaximalentanglementEAQAG}, such that
 \begin{equation}
  R_{\mathcal{Q}_1}  =\frac{K_1}{N_1}  \geq 1-m_1\delta_1-\frac{1}{A(2^{m_1})},
 \end{equation}
 where $0\leq\delta_1\leq(1 -1/A(2^{m_1}))/m_1$ is the relative minimum distance of $\mathcal{Q}_1$.

 If $m_2>2$ is odd, then there exists an $[[m_2,m_2-1,2;1]]_2$ binary EAQECC with maximal entanglement according to Refs.  \cite{lai2017linear,brun2014catalytic}. We take the inner code as $Q_I=[[m_2,m_2-1,2;1]]_2$ and  take the outer code as the EAQAG code with parameters $Q_O=[[N_2, K_2,D_2;N_2-K_2]]_{2^{m_2-1}}$.  Then there exists an EACQC with parameters
$
 \mathcal{Q}_2 =[[m_2N_2,(m_2-1)K_2,2D_2;m_2(N_2-K_2)+K_2]]_2
$ from Theorem \ref{EACQC scheme}. We can derive the asymptotic bound of $
 \mathcal{Q}_2$ from Lemma \ref{asymptoticmaximalentanglementEAQAG}, such that
 \begin{eqnarray}
 R_{\mathcal{Q}_2}  &=&\frac{(m_2-1)K_2}{m_2N_2}  \\
 &\geq&   \left(1-\frac{1}{m_2}\right)\left(1-\frac{m_2}{2}\delta_2-\frac{1}{A(2^{(m_2-1)})}\right),
 \end{eqnarray}
 where $0\leq\delta_2\leq 2(1 -1/A(2^{(m_2-1)}))/m_2$ is the relative minimum distance of $\mathcal{Q}_2$.
\end{IEEEproof}

In Lemma \ref{asymptoticmaximalentanglementEAQAG}, if we take $q$ as a square of a prime, then $A(q)$ can achieve the maximum number of rational points  of  a curve   $\mathcal{X}/\mathbb{F}_q$ of genus $g$. In such case, we can take $A(q)=\sqrt{q}-1$ explicitly.
\begin{corollarys}
There exists a family of  binary EACQCs $\mathcal{Q}$   consuming maximal entanglement   with asymptotic parameters such that
 \begin{equation}
\label{asymptoticGVm1}
 R_{\mathcal{Q}}  \geq 1-m\delta-\frac{1}{ 2^{m/2}-1 },
  \end{equation}
 where $m>1$ is even, $R_{\mathcal{Q}}$ is the   rate of $\mathcal{Q}$ and $0\leq\delta\leq(1 -1/(2^{m/2}-1))/m$ is the relative minimum distance of $\mathcal{Q}$.
\end{corollarys}
 \begin{corollarys}
There exists a family of  binary EACQCs $\mathcal{Q}$   consuming maximal entanglement   with asymptotic parameters such that
 \begin{equation}
\label{asymptoticGVm2}
  R_{\mathcal{Q} }    \geq  \left (1-\frac{1}{m }\right)\left(1-\frac{m }{2}\delta -\frac{1}{2^{(m-1)/2}-1}\right),
  \end{equation}
 where $m>1$ is odd, $R_{\mathcal{Q}}$ is the   rate of $\mathcal{Q}$ and $0\leq\delta\leq 2(1 -1/(2^{(m-1)/2}-1)/m$ is the relative minimum distance of $\mathcal{Q}$.
\end{corollarys}

It is known that the entanglement-assisted quantum codes need to consume extra entangled states during the transmission. Although the extra  entanglement  can boost the transmission ability  of EAQECCs, it does not make sense when we compare with standard QECCs \cite{brun2014catalytic}. Therefore we usually take EAQECCs with positive net transmissions  for many purposes, e.g., catalytic quantum error correction and comparisons of EAQECCs with standard QECCs. If we only  consider     EACQCs with positive  net transmissions  in Proposition \ref{propositionmaxEACQC}, then we have the following two results.
\begin{corollarys}
\label{asymptoticGVm1positive}
There exists a family of  asymptotically good binary EACQCs $\mathcal{Q}$   consuming maximal entanglement   with positive net transmissions  such that
 \begin{equation}
 R_{\mathcal{Q}}  \geq 1-2m\delta-\frac{2}{ 2^{m/2}-1 },
  \end{equation}
 where $m>3$ is even, $R_{\mathcal{Q}}$ is the   rate of $\mathcal{Q}$ and $0\leq\delta<(1 -2/(2^{m/2}-1))/(2m)$ is the relative minimum distance of $\mathcal{Q}$.
\end{corollarys}
 \begin{corollarys}
 \label{asymptoticGVm2positive}
There exists a family of   asymptotically good binary EACQCs $\mathcal{Q}$   consuming maximal entanglement   with positive net transmissions  such that
 \begin{equation}
  R_{\mathcal{Q} }    \geq   \left(2-\frac{2}{m}\right)\left(1-\frac{m }{2}\delta -\frac{1}{2^{(m-1)/2}-1}\right)-1,
  \end{equation}
 where $m>4$ is odd, $R_{\mathcal{Q}}$ is the   rate of $\mathcal{Q}$ and $0\leq\delta\leq  (1 -1/(m-1)-2/(2^{(m-1)/2}-1))/m$ is the relative minimum distance of $\mathcal{Q}$.
\end{corollarys}

\begin{figure}[!h]
    \centering
    \includegraphics[width=1.0\linewidth]{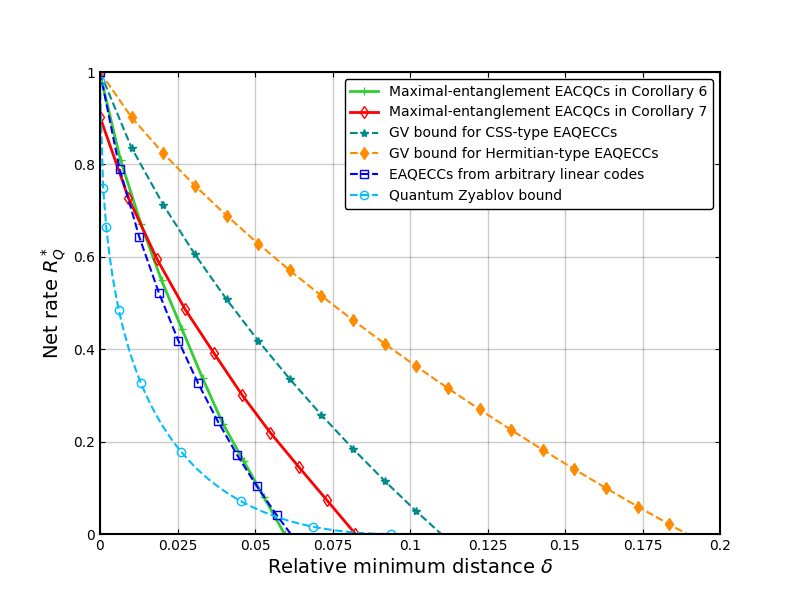}
    \caption{The asymptotic performance of EACQCs with maximal entanglement.}
    \label{Asymptotic boundss}
\end{figure}

In Fig. \ref{Asymptotic boundss}, we plot the asymptotic bounds of several EACQCs and EAQECCs. We also plot the famous GV bound for EAQECCs and the quantum Zyablov bound. In \cite{qian2015entanglement}, asymptotically good EAQECCs with maximal entanglement were constructed from arbitrary linear codes and we also plot the asymptotic bound in \cite{qian2015entanglement}. It is shown in Fig. \ref{Asymptotic boundss} that the asymptotic bounds in Corollary \ref{asymptoticGVm1positive} and Corollary \ref{asymptoticGVm2positive} can    outperform the quantum Zyablov bound and the asymptotic bound of \cite{qian2015entanglement}.

\section{Asymptotically Good EACQCs   Meeting  the   Quantum GV Bound For EAQECCs}\label{AsymptoticallygoodEACQCs}
It was shown in \cite{blokh1973existence} that    classical concatenated codes  can attain the GV bound   asymptotically  by selecting   the inner and outer codes  systematically and    randomly.   Then in \cite{thommesen1983existence}, by fixing the outer codes on Reed-Solomon codes and  varying  the nonsystematic inner codes  randomly,   the resultant  concatenated codes can also attain the GV bound   asymptotically. In \cite{ouyang2014concatenated}, the idea in \cite{thommesen1983existence} was generalized to CQCs and it was shown that CQCs can attain the quantum GV bound  asymptotically. In this section, we generalize the idea in \cite{blokh1973existence} from classical concatenated codes to EACQCs. We show that EACQCs can also attain the GV bound for EAQECCs asymptotically if we select  both the inner and outer EAQECCs randomly.

Let $C_1=[n_1,k_1,d_1]_4$ be a quaternary code whose generator and parity-check matrices are given by $G_1$ and $H_1$, respectively. Then there exists a binary EAQECC with parameters $\mathscr{Q}_1=[[n_1,2k_1-n_1+c_1,d_1;c_1]]_2$ where $c_1=rank(H_1H_1^\dagger)$. Denote by $r_1=n_1-k_1$. Denote by $\bar{\textbf{k}}_1=2k_1-n_1+c_1$ and let $C_2=[n_2,k_2,d_2]_{4^{\bar{\textbf{k}}_1}}$ be a classical $ {4^{\bar{\textbf{k}}_1}}$-ary linear code  with the generator and parity-check matrices   given by $G_2$ and $H_2$, respectively. Denote by $r_2=n_2-k_2$. There exists an EAQECC with parameters $\mathscr{Q}_2=[[n_2,\bar{\textbf{k}}_2,d_2;c_2]]_{2^{\bar{\textbf{k}}_1}}$ where      $c_2=rank(H_2H_2^\dagger)$ and $\bar{\textbf{k}}_2=2k_2-n_2+c_2$. Let $\beta=\{\alpha_1,\cdots,\alpha_{\bar{\textbf{k}}_1}\}$ be a basis of $\mathbb{F}_{4^{\bar{\textbf{k}}_1}}$. Then there is a one to one correspondence between $\mathbb{F}_{4^{\bar{\textbf{k}}_1}}$  and $\mathbb{F}_{4}^{\bar{\textbf{k}}_1}$ under the basis $\beta$. We denote by $\widetilde{u}\leftrightarrow u$ for   $u\in \mathbb{F}_{4^{\bar{\textbf{k}}_1}}$ and $\widetilde{u}\in \mathbb{F}_{4}^{\bar{\textbf{k}}_1}$.
  For simplicity,  we do not consider degenerate errors for EAQECCs $\mathscr{Q}_1$ and $\mathscr{Q}_2$ here. For classical linear codes $C_1$ and $C_2$, we suppose that they are both system  codes, i.e., $G_1=[I\ P_1]$ and $G_2=[I\ P_2]$ are in systematic formats. Moreover, suppose that   $P_1$ and $P_2$  are both random matrices and the elements of them  are both generated uniformly. According to Lemma \ref{EACQC scheme}, there exists an EACQC with parameters
\begin{equation}
\mathscr{Q}_e=[[n_e=n_1n_2,\bar{\textbf{k}}_1\bar{\textbf{k}}_2,d_e\geq d_1d_2;c_e]]_2,
\end{equation}
where $c_e= c_1n_2+c_2\bar{\textbf{k}}_1$. Denote by $r_e=n_e-\bar{\textbf{k}}_1\bar{\textbf{k}}_2$ and denote  $R_e=\bar{\textbf{k}}_1\bar{\textbf{k}}_2/n_e$ by the rate of $\mathscr{Q}_e$. Denote $C_e=c_e/n_e$ by the entanglement-assistance rate of  $\mathscr{Q}_e$. If $R_e>C_e$, then we say that $\mathscr{Q}_e$ has positive transmissions and we denote $R_e^*=R_e-C_e$  by the net transmission rate.

 We recall the encoding process of EACQCs in \cite{fan2022entanglement}.
First we  encode the quantum information state $|\alpha\rangle$ by using the outer code $\mathscr{Q}_2$ as follows
 \begin{equation}
 |\alpha\rangle  \mapsto |\varphi\rangle_O= (U_{O}\otimes \widehat{I}_{B_O})|\alpha\rangle\otimes|0\rangle^{\otimes
(n_2-\bar{\textbf{k}}_2-c_2) \bar{\textbf{k}}_1}\otimes |\Psi_+\rangle_{AB}^{c_2\bar{\textbf{k}}_1}.
 \end{equation}
where $|\Psi_+\rangle_{AB}^{c_2\bar{\textbf{k}}_1}$   is the Bell state preshared between Alice and
 Bob  during the outer encoding. The  operation $U_O$ is applied to the qubits in Alice's side.
 Bob's halves of quantum states are preshared before the time and they do not need to be encoded.
We   represent the quantum state $|\varphi\rangle_O $ by
\begin{equation}
    |\varphi\rangle_O =\sum_{\mu_1,\cdots,\mu_{n_2}=0}^{2^{\bar{\textbf{k}}_1}}a_{\mu_1\cdots \mu_{n_2}}|\mu_1\cdots\mu_{n_2}\rangle,
\end{equation}
where the coefficients $a_{\mu_1\cdots \mu_{n_2}} (0\leq\mu_1,\cdots,\mu_{n_2}\leq 2^{\bar{\textbf{k}}_1} )$ should satisfy  the normalization condition.
Then we separate each basis   $|\mu_1\cdots\mu_{n_2}\rangle$   into $n_2$ subblocks, i.e., $|\mu_1\cdots\mu_{n_2}\rangle=|\mu_1\rangle\cdots|\mu_{n_2}\rangle$ for    $0\leq\mu_1,\cdots,\mu_{n_2}\leq 2^{\bar{\textbf{k}}_1}$.
  For each subblock $| \mu_i\rangle (1\leq i\leq n_2)$,  we do the inner encoding as follows:
\begin{equation}
| \mu_i\rangle \mapsto |\psi_i\rangle_I= (U_{I}\otimes \widehat{I}_{B_I})| \mu_i\rangle\otimes|0\rangle^{\otimes n_1-\bar{\textbf{k}}_1-c_1}\otimes |\Phi_+\rangle_{AB}^{c_1}.
 \end{equation}
  $|\Phi_+\rangle_{AB}^{c_1}$   are ${c_1 }$ Bell states preshared between Alice and
Bob during each inner   encoding. The inner encoding operation $U_I$ is applied to the qubits in Alice's side
while Bob's halves of ebits  do not need to be encoded.

 Let $\textbf{u} $ be a quaternary vector of length $n_e=n_1n_2$ with elements over $\mathbb{F}_4$. We divide it into $n_2$ subblocks  and each subblock has a length of $n_1$, i.e., $\textbf{u}=(\textbf{u}_1,\cdots,\textbf{u}_{n_2})$ and $\textbf{u}_i=(u_{i1},\cdots,u_{in_1})$ with $u_{ij}\in \mathbb{F}_4$ for $1\leq i\leq n_2$ and $1\leq j\leq n_1$. Suppose that $\textbf{u} $ has a weight $w$ and suppose that there are $t$   subblocks in $\textbf{u} $ such that each of them contains at least one non-zero element and the other $n_2- t$ subblocks are all zeros.   We impose $\textbf{u} $ as a quantum error to the codeword of the EACQC $\mathscr{Q}_e$. For each inner code $\mathscr{Q}_1$ and each  $\textbf{u}_i (1\leq i\leq n_2)$, the syndrome information is given as $S_{1i}=H_1\textbf{u}_i^T(1\leq i\leq n_2)$. If $S_{1i}=0$ for some $1\leq i\leq n_2$, then $\textbf{u}_i$ is a codeword $C_1$. For each $ \textbf{u}_i (1\leq i\leq n_2)$,  we denote by $ \textbf{u}_i =(\overline{u}_i,\overline{v}_i)$ where $\overline{u}_i$ is of length $k_1$ and $\overline{v}_i$ is of length $n_1-k_1$. If some $\textbf{u}_i  $ is a zero vector, then the probability that $S_{1i}=0$ is definitely $1$. Thus we need to consider the case that $\textbf{u}_i  $ is a nonzero vector. If $\overline{u}_i=0$ for some nonzero $\textbf{u}_i  $, then the   probability that $S_{1i}=0$ is $0$ since the generator and parity-check matrices of $C_1$ are systematic. If  $\overline{u}_i\ne0$ for some nonzero $\textbf{u}_i  $,  then the   probability that $S_{1i}=0$ is $4^{-r_1}$. Notice that this probability is irrelevant to the weight of $\textbf{u}_i  $. Such property greatly simplifies our counting since we only need to consider wether $\textbf{u}_i  $ is zero or not.

 According to the encoding circuit of quantum codes, there is a one to one correspondence between the information   states and the encoded   states since the encoding operation is unitary. Therefore if we run backward the encoding circuit of each inner code, then there are exactly $t$ errors imposed on the outer code. Let $\textbf{u} $  be a map of $\widetilde{\textbf{u}}$ under a basis $\beta $ of $\mathbb{F}_{4^{\bar{\textbf{k}}_1}}$. Denote by $S_2=H_2 \widetilde{\textbf{u}}^T$.  Then the probability that $S_{2}=0$ is either $0$ or $(4^{\bar{\textbf{k}}_1})^{-r_2}=4^{-\bar{\textbf{k}}_1r_2}$.  Denote  $P_t(w)$ by the probability that  the
quaternary vector $\textbf{u} $   with weight $w>0$ such that it is a codeword of the stabilizer code of  a
randomly chosen EACQC. Then we have $P_t(w)$ is either $0$ or equal to $4^{-tr_1-\bar{\textbf{k}}_1r_2}$. Then we have the following result about EACQCs.
\begin{lemmas}\label{ntwPsix}
 Let $\mathcal{N}_t(w)$ be the number of all quaternary vectors $\textbf{u}$ of length $n_e=n_1n_2$ and weight $w$ that contain exactly $t$ nonzero
subblocks and $n_2-t$ zero subblocks. Denote
\begin{equation}
\Psi_t(x)=\sum_{w=t}^{n_e}\mathcal{N}_t(w)x^w
\end{equation}
by the generating function for the quantity $\mathcal{N}_t(w)$. Then there is
\begin{equation}
\Psi_t(x) = \binom{n_2}{t}[(1+3x)^{n_1}-1]^t
\end{equation}
 \end{lemmas}
 \begin{IEEEproof}
 There exists the following relationship to $\mathcal{N}_t(w)$ such that \begin{equation}\mathcal{N}_t(w)=\binom{n_2}{t}M_t(w)\end{equation}
 where $M_t(w)$ is the number of quaternary vectors  $v=(v_1,\cdots,v_t)$ whose elements    are non-zero values. Then the quantity $M_t(w)$ satisfies the following recursion relationship
 \begin{equation}
M_t(w) =  \sum_{i=1}^{n_1}3^i\binom{n_1}{i}M_{t-1}(w-i)
\end{equation}
where $M_t(w) =0  $ if $w<t$. Denote by $\chi_t(x)= \sum_{w=t}M_t(w)x^w$. Then there is
\begin{eqnarray}
\chi_t(x) &=& \chi_{t-1}(x) \sum_{i=1}^{n_1} 3^i\binom{n_1}{i}x^i \nonumber\\
&=&\chi_{t-1}(x)[(1+3x)^{n_1}-1] \nonumber\\
&= &\chi_1(x)[(1+3x)^{n_1}-1]^{t-1}.
\end{eqnarray}
Observing that \begin{eqnarray}
\chi_1(x)&=&\sum_{w=1}^{n_1}M_1(w)x^w \nonumber\\
&=&\sum_{w=1}^{n_1}3^w\binom{n_1}{w}x^w \nonumber\\
&=&(1+3x)^{n_1}-1 ,
\end{eqnarray}
 then we have $\chi_t(x) = [(1+3x)^{n_1}-1]^t$ and thus $\Psi_t(x)= \binom{n_2}{t}[(1+3x)^{n_1}-1]^t$.
 \end{IEEEproof}

 Let $ \overline{N}(w)$ be the average number of codewords of
weight $w > 0$ in the stabilizer code of an EACQC  over the entire
ensemble of random EACQCs. Then
\begin{eqnarray}
 \overline{N}(w) & = &\sum_{t=1}^{n_2} \mathcal{N}_t(w)P_t(w)\nonumber\\ \label{averagenocdws}
 &\leq & \sum_{t=1}^{n_2}  \mathcal{N}_t(w)4^{-tr_1-\bar{\textbf{k}}_1r_2}
 \end{eqnarray}
 since  $P_t(w)$ is either $0$ or equal to $4^{-tr_1-\bar{\textbf{k}}_1r_2}$ when $w>0$.
 \begin{lemmas}\label{varphixNwxw}
 Denote the generating function
 \begin{equation}\label{nwxvarphix}
 \varphi(x) = \sum_{w=1}^{n_e} \overline{N}(w)x^w
 \end{equation}
 by the average number of codewords of weight $w>0$ across
the ensemble of the stabilizer code of EACQCs. There exists
\begin{equation}
 \varphi(x) \leq 2^{-r_e-c_e}[(1+3x)^{n_1}+4^{r_1}]^{n_2}.
\end{equation}
\end{lemmas}
\begin{IEEEproof}
According to Eq. (\ref{averagenocdws}) and Lemma \ref{ntwPsix}, there is
\begin{eqnarray}
\varphi(x) &\leq&\sum_{w=1}^{n_e}\sum_{t=1}^{n_2}  \mathcal{N}_t(w)4^{-tr_1-\bar{\textbf{k}}_1r_2}x^w \nonumber\\
 &=&\sum_{t=1}^{n_2}4^{-tr_1-\bar{\textbf{k}}_1r_2} \binom{n_2}{t}[(1+3x)^{n_1}-1]^t \nonumber\\
 &=&4^{-\bar{\textbf{k}}_1r_2} \left[\frac{(1+3x)^{n_1}-1}{4^{r_1}}+1\right]^{n_2} \nonumber \\
 &\leq&4^{-\bar{\textbf{k}}_1r_2-r_1n_2}[(1+3x)^{n_1}+4^{r_1}]^{n_2}\nonumber \\
 &=&2^{-r_e-c_e}[(1+3x)^{n_1}+4^{r_1}]^{n_2}.
 \end{eqnarray}
\end{IEEEproof}
  \begin{lemmas}\label{verage number of codewordsofnw}
  The average number of codewords $ \overline{N}(w)$ of the stabilizer code of EACQCs
  with weight $w > 0$ across the ensemble
satisfies the following inequality
  \begin{equation}
\overline{N}(w)\leq2^{ {n_e}(R_e+2H_4(\gamma)-1-C_e)}[1+4^{r_1}(1-\gamma)^{n_1}]^{n_2},
  \end{equation}
  where $w=\gamma n_e$ and $H_4(\gamma)= \gamma\log_4 3-\gamma\log_4\gamma-(1-\gamma)\log_4(1-\gamma)$ is the quaternary entropy function.
  \end{lemmas}
 \begin{IEEEproof}
 It follows from Eq. (\ref{nwxvarphix}) in  Lemma \ref{varphixNwxw} that $ \overline{N}(w) x^w\leq \varphi(x)$ which implies that
 \begin{eqnarray}
  \overline{N}(w) &\leq& x^{-w}\varphi(x) \nonumber \\
  &\leq& x^{-w}2^{-r_e-c_e}[(1+3x)^{n_1}+4^{r_1}]^{n_2}\nonumber\\
  &\leq& 2^{-r_e-c_e}\left[ (x^{-\gamma}+3x^{1-\gamma})^{n_1} + 4^{r_1}x^{-\gamma n_1}   \right]^{n_2}.
 \end{eqnarray}
 Compute the derivative  of $x^{-\gamma}+3x^{1-\gamma}$ and   take $ x=\gamma/(3-3\gamma) $. Then $x^{-\gamma}+3x^{1-\gamma}$ can achieve its minimum value. Considering that
 \begin{equation}
\left\{
             \begin{aligned}
              \left(\frac{3(1-\gamma)}{\gamma}\right)^{\gamma} +3\left(\frac{3(1-\gamma)}{\gamma}\right)^{\gamma-1}
           &=3(3(1-\gamma))^{\gamma-1}\gamma^{-\gamma}
\\
&=4^{H_4(\gamma)},\\
          (3(1-\gamma))^{\gamma}\gamma^{-\gamma}4^{-H_4(\gamma)}&=1-\gamma.
             \end{aligned}
\right.
\end{equation}

  Thus we have
  \begin{equation}
\overline{N}(w)\leq2^{ n_e(R_e+2H_4(\gamma)-1 -C_e) }[1+4^{r_1}(1-\gamma)^{n_1}]^{n_2}.
  \end{equation}
 \end{IEEEproof}

Denote by $\mathcal{L}(w)=\sum_{i=1}^{w}\overline{N}(i)$ and denote by $\rho(x)=\sum_{w=1}^{n_e}\mathcal{L}(w)x^w$. Denote $P(d_e\leq \delta_en_e)$ by   the probability
 that the stabilizer code of a random EACQC has a distance $d_e\leq \delta_en_e$. There exists the following result about EACQCs.
\begin{theorems}
The probability $\mathcal{P}=P(d_e\leq \delta_en_e)$ satisfies
\begin{equation}
\mathcal{P}\leq \frac{1- \delta_e}{1-2\delta_e}2^{{n_e}(R_e+2H_4(\delta_e)-1-C_e)}[1+4^{r_1}(1-\delta_e)^{n_1}]^{n_2}.
\end{equation}
There exists a family of  asymptotically good EACQCs such that $\mathcal{P}\rightarrow 0$ and $
R_e\rightarrow 1+C_e-2H_4(\delta_e)
$ as $n_e\rightarrow\infty$.
\end{theorems}
\begin{IEEEproof}
Since $\overline{N}(i)=0$ and $\mathcal{L}(i)=\mathcal{L}(n_e)$ for all $i>n_e$. Then for $0<x<1$, there is
\begin{equation}
\rho(x) = \sum_{i=1}^\infty \mathcal{L}(i)x^i-\mathcal{L}(n_e)x^{n_e+1}(1-x)^{-1}\leq  \sum_{i=1}^\infty\mathcal{L}(i)x^i,
\end{equation}
where \begin{equation}
\sum_{i=1}^\infty \mathcal{L}(i)x^i = (1-x)^{-1}\sum_{i=1}^{n_e}\overline{N}(i)x^i=(1-x)^{-1}\varphi(x).
\end{equation}
Therefore \begin{equation}
 \rho(x) \leq 2^{-r_e-c_e}(1-x)^{-1}[(1+3x)^{n_1}+4^{r_1}]^{n_2}.
\end{equation}
Notice that
\begin{equation}
P(d_e\leq \delta_en_e)\leq \sum_{i=1}^{\delta_en_e}\overline{N}(i)=\mathcal{L}(\delta_en_e)\leq x^{-\delta_en_e} \rho(x)
\end{equation}
for all $0<x<1$. Let $x=\delta_e/(1-\delta_e)$. Then according to Lemma \ref{verage number of codewordsofnw}, there is
\begin{equation}
\mathcal{P} \leq \frac{1- \delta_e}{1-2\delta_e}2^{{n_e}(R_e+2H_4(\delta_e)-1-C_e)}[1+4^{r_1}(1-\delta_e)^{n_1}]^{n_2}.
\end{equation}
 Denote by $x_0$ the   solution to the
equation $2H_4(x)=1-R_e+C_e$. We set $\delta_e=x_0-\epsilon$ where $\epsilon>0$ is a sufficiently small constant. Let the transmission rate of the inner code $C_1$ satisfy the condition  such that
\begin{equation}
0<\tau\equiv4^{1-R_1}(1-\delta_e)=4^{r_1/n_1}(1-\delta_e)<1 .
\end{equation}
The code length $n_2$ of the outer code $C_2$ should be chosen such that $c=\tau^{n_1}n_2 > 0$ is a constant. Thus there is
\begin{equation}
[1+4^{r_1}(1-\delta_e)^{n_1}]^{n_2} =(1+\tau^{n_1})^{n_2}=(1+\frac{c}{n_2})^{n_2}\leq e^c.
 \end{equation}
Moreover,  there is $2^{ {n_e}(R_e+2H_4(\delta_e)-1-C_e)}\rightarrow 0$ as $n_e=n_1n_2\rightarrow\infty$ for arbitrary $\delta_e<x_0$. Thus we have $\mathcal{P}=P(d_e\leq \delta_en_e)\rightarrow 0$ as $n_e \rightarrow \infty$. Therefore, to conclude, there exist EACQCs with a probability of $1$  as $n_e \rightarrow \infty$ such that the code distance $d_e\geq  \delta_en_e$ and the rate $R_e=1+C_e-2H_4(\delta_e)$.  Thus EACQCs satisfy the quantum GV bound for EAQECCs asymptotically.
\end{IEEEproof}

 \section{Conclusion and Discussion}
\label{ConclusionandDiscussion}
 This paper is devoted to the construction of   EACQCs with new  parameters and the asymptotic performance. We used EAQAMDS, $\hbar_e$-EAQMDS   and EAQAG codes  as the outer codes to construct   new EACQCs. A large amount of EACQCs with parameters beating the previously best known  QECCs and EAQECCs of the same length and net transmissions were derived. We showed that  EACQCs are with maximal
 entanglement if both the inner and outer   codes are with maximal entanglement. We derived three new    optimal  EACQCs with maximal entanglement. We  also derived several new maximal-entanglement EACQCs whose minimum distance is only one less than the minimum distance of the optimal codes.  In particular,   we  explicitly  constructed two new families of asymptotically good EACQCs with maximal entanglement by using  EAQAG codes  as the outer codes.  At last, we proved that  EACQCs    can attain the quantum GV bound for EAQECCs asymptotically. In the future work, how to apply the idea of EACQCs to fault-tolerant quantum computations and quantum communications is an interesting problem.


\ifCLASSOPTIONcaptionsoff
  \newpage
\fi



%
\bibliographystyle{IEEEtran}
\bibliography{IEEEabrv,thebibfile}

\end{document}